\documentclass[aps,prl,twocolumn,longbibliography,superscriptaddress]{revtex4-1}
\usepackage{amsmath,amssymb,mathrsfs}
\usepackage{natbib}
\usepackage{subfigure}
\usepackage{tabularx}
\usepackage{epsfig}
\usepackage{longtable}
\usepackage{amsfonts}
\usepackage{rotating}
\usepackage{subfigure}
\usepackage{amsmath}
\usepackage{comment}
\usepackage{bbold} 
\usepackage{multirow}
\usepackage{hhline}

\def\be{\begin{equation}}
\def\ee{\end{equation}}
\def\bea{\begin{eqnarray}}
\def\eea{\end{eqnarray}}
\def\nn{\nonumber}

\usepackage[unicode=true,bookmarks=true,bookmarksnumbered=false,bookmarksopen=false,breaklinks=false,pdfborder={0 0 1},backref=false,colorlinks=true]{hyperref}

\hypersetup{linkcolor=magenta,urlcolor=blue,citecolor=blue,pdfstartview={FitH},hyperfootnotes=false,unicode=true}

\def\blue{\color{blue}}

\def\be{\begin{equation}}
\def\ee{\end{equation}}
\def\bea{\begin{eqnarray}}
\def\eea{\end{eqnarray}}
\def\nn{\nonumber}

\begin{document}
\title{Atom-Orbital  Qubit under Non-Adiabatic Holonomic Quantum Control}
\author{Hongmian Shui$^{\blue \ddagger}$} 
\affiliation{State Key Laboratory of Advanced Optical Communication System and Network, Department of Electronics, Peking University, Beijing 100871, China} 
\author{Shengjie Jin$^{\blue \ddagger}$} 
\affiliation{State Key Laboratory of Advanced Optical Communication System and Network, Department of Electronics, Peking University, Beijing 100871, China} 
\author{Zhihan Li$^{\blue \ddagger}$} 
\affiliation{State Key Laboratory of Advanced Optical Communication System and Network, Department of Electronics, Peking University, Beijing 100871, China} 
\author{Fansu Wei} 
\affiliation{State Key Laboratory of Advanced Optical Communication System and Network, Department of Electronics, Peking University, Beijing 100871, China} 
\author{Xuzong Chen} 
\affiliation{State Key Laboratory of Advanced Optical Communication System and Network, Department of Electronics, Peking University, Beijing 100871, China} 
\author{Xiaopeng Li}  
\email{xiaopeng\_li@fudan.edu.cn}
\affiliation{State Key Laboratory of Surface Physics, Institute of Nanoelectronics and Quantum Computing, and Department of Physics, Fudan University, Shanghai 200438, China}
\affiliation{Shanghai Qi Zhi Institute, Shanghai 200030, China}
\author{Xiaoji Zhou} 
\email{xjzhou@pku.edu.cn}
\affiliation{State Key Laboratory of Advanced Optical Communication System and Network, Department of Electronics, Peking University, Beijing 100871, China} 
\affiliation{Collaborative Innovation Center of Extreme Optics, Shanxi University, Taiyuan, Shanxi 030006, China}

\begin{abstract}
Quantum computing has been attracting tremendous efforts in recent years. One prominent application is to perform programmable quantum simulations of electron correlations in large molecules and solid-state materials, where orbital degrees of freedom are crucial to quantitatively model electronic properties.  Electron orbitals unlike quantum spins obey crystal symmetries, making the atomic orbital in optical lattices  a natural candidate to emulate electron orbitals.  
Here, we construct an atom-orbital {qubit} by manipulating $s$- and $d$-orbitals of atomic Bose-Einstein condensation in an optical lattice.  
Noise-resilient single-qubit gates are achieved by performing holonomic quantum control, which admits geometrical protection. We find  it is critical to eliminate the orbital leakage error in the system.  Our work opens up wide opportunities for atom-orbital based quantum information processing, of vital importance to programmable quantum simulations of multi-orbital physics in molecules and quantum materials.

\end{abstract}

\date{\today}


\maketitle


Orbital degrees of freedom are essential to the quantitative description of electrons in large molecules~\cite{2007_Bartlett_RMP} and solid state materials~\cite{2012_Hwang_NatMat}.  
The complex interplay of spin, charge, and orbital, is key to the emergence of novel electron phenomena such as multiferroics~\cite{2012_Hwang_NatMat}, unconventional superconductivity~\cite{2011_Stewart_RMP}, and exotic molecular spin filtering~\cite{2020_Li_Chiral,2021_Yan_NatMat}. Incorporating orbitals lies at the heart of accurate quantum chemistry calculations~\cite{2007_Bartlett_RMP}, and adds substantial computation complexity in simulating many-body electron correlation. With quantum computing, the overall computation complexity of quantum algorithms for quantum chemistry calculation has been reduced to polynomial~\cite{aspuru2005simulated,2020_McArdle_RMP}, which has triggered much recent research effort on experimental demonstration of such quantum algorithms with superconducting qubits~\cite{kandala2017hardware,google2020hartree} and trapped ions~\cite{2018_Blatt_PRX,nam2020ground}. However,  using these qubits to emulate electron orbitals meets experimental challenge of expensive qubit encoding~\cite{kandala2017hardware} and demanding Hamiltonian engineering  to impose precise orbital symmetry~\cite{2016_McClean_NJP}, and consequently the captured multi-orbital effect in the experiments is rather limited~\cite{kandala2017hardware,google2020hartree}.

With ultracold atoms confined in optical lattices, the atom-orbital wavefunction obeys the same crystalline symmetry as the electron orbitals~\cite{Li_2016}, making atom-orbital an ideal qubit candidate to perform programmable quantum simulations of electron orbitals in molecules and solid  materials.
There has been fascinating progress accomplished in controlling atom orbitals in optical lattices for quantum simulations of exotic superfluids~\cite{2005_Girvin_PRA,2006_Liu_PRA,2006_Kuklov_PRL,2011_Wirth_NatPhys,2012_Sengstock_NatPhys,Jin_PRL2021} and topological quantum states~\cite{2012_Sun_NatPhys,2013_Li_NC,2014_Liu_NC,2016_Engels_NC}.  
However, their gate-based universal quantum  control is still lacking, leaving orbital qubit quantum computing so-far unachieved.

\begin{figure}[htp]
\centering
\includegraphics[width=.48\textwidth]{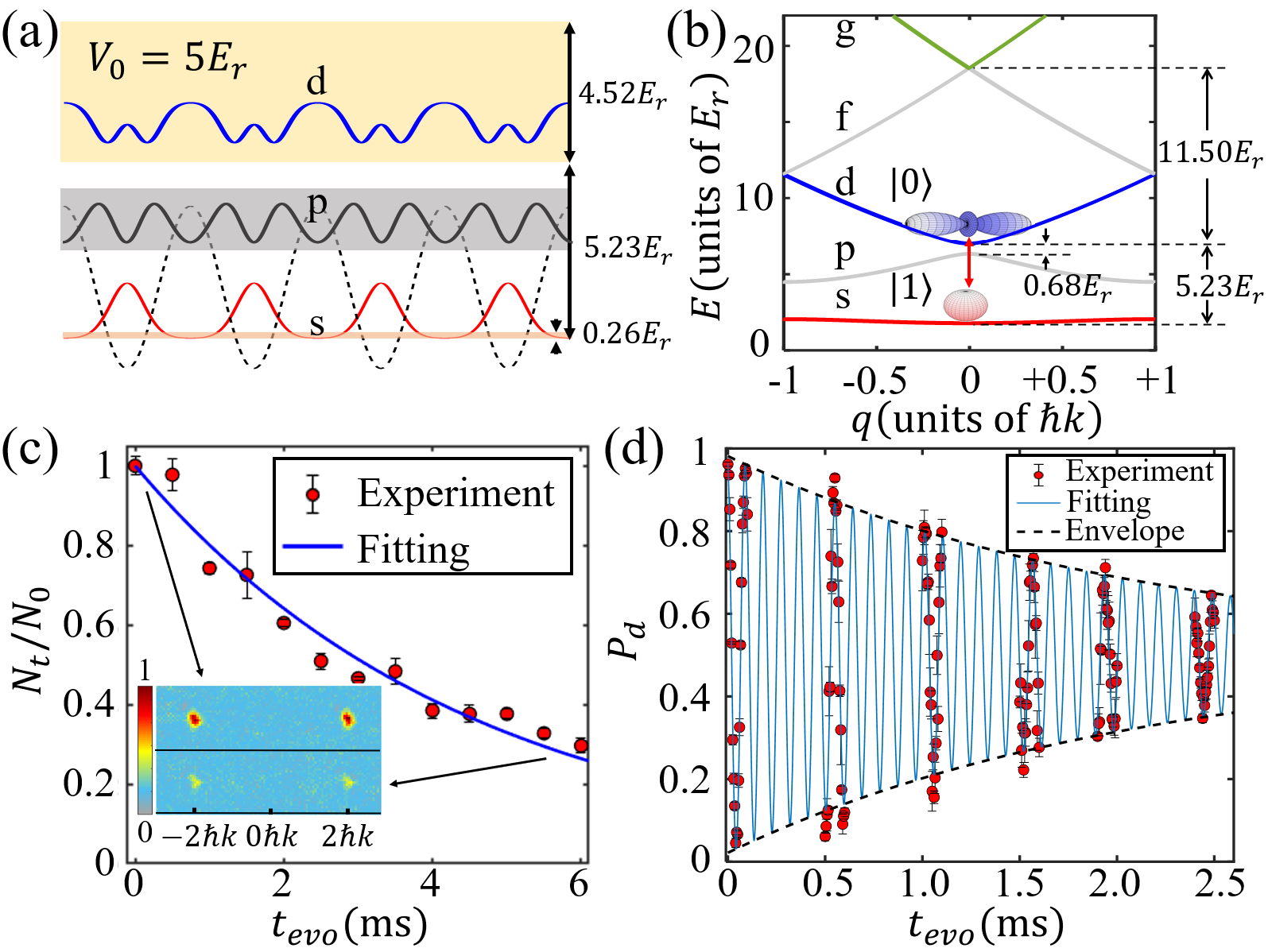}
\caption{Atom orbital {qubit} in a one-dimensional optical lattice. (a) The Bloch wave function amplitude of the $s$, $p$, and $d$ bands at zero quasi-momentum. The $s$- and $d$- states constitute our orbital {qubit}. The dashed line represents the optical lattice potential. 
(b) The band gap structures in our experimental optical lattice. The energy bands from bottom to top are $s$, $p$, $d$, $f$ and $g$ bands respectively.
(c) Measurement of relaxation time, $T_{\rm relax} $, in the decay dynamics of $d$-band population. The red dots correspond to  the time evolution of the atom number $N_d$ in  the $d$ band by time-of-flight measurement. The blue line represents fitting to a function $N_d(t)=N_0\exp(-t/T_{\rm relax} )$. The relaxation time is determined as $T_{\rm relax} =4.5$ ms.
(d) Measurement of dephasing time, $T_2$, through Ramsey interference. The red dots show the measured time evolution of  the proportion of atoms $P_{d}$ in the $d$-band by band mapping method. The blue line represents a fitting function $P_{d}=A\exp(-t/T_2)\sin(\omega t+\phi)+B$, with  the black line the envelope of the Ramsey oscillations. We  have a dephasing time $T_2=2.1$ ms by fitting.}
\label{fig:orqbit}
\end{figure}

In this Letter, we  construct an atom-orbital qubit using $s$- and $d$-orbitals of a one-dimensional optical lattice. We have measured the orbital relaxation  time ($T_{\rm relax} $) and the dephasing  time ($T_2$), finding {$T_{\rm relax} = 4.5\pm0.1$ ms (milliseconds), and $T_2 = 2.1\pm0.1$ ms} in our experiment. By programming  lattice modulation, we reach universal  nonadiabatic holonomic quantum gate control~\cite{1969_LewisJMP,2014_Gungordu_JPSJ} over the atom-orbital qubit, which exhibits noise-resilience against laser fluctuations due to geometrical protection. We demonstrate the holonomic quantum control of Hadamard and $\pi/8$ gates, which form a universal gate-set for single-qubit rotation.  The lattice modulation pulses are programmed to minimize orbital leakage error, 
{which is the key to reach high fidelity holonomic quantum control of atom-orbital {qubit}.} 
We implement quantum process tomography (QPT) on the orbital {qubit} to measure the full density matrix, from which  the obtained average gate fidelity is $98.36(10)\%$.


{\it Atom-orbital qubit.---}
Our experiment is based on a $^{87}$Rb Bose-Einstein condensate (BEC) with $2\times 10^5$ atoms confined in a one-dimensional (1D) optical lattice.  The lattice potential takes a form of $V_{p}(x) = V_0 \cos^2 (2\pi x/l)$, with $V_0$ the lattice depth, and $l$ the wavelength of the laser forming the lattice, which is $1064$ nm in this experiment.  The lattice depth $V_0$ is five times of the laser recoil energy ($E_r=h^2/2ml^2$), with $m$ the atomic mass.  
With lattice confinement, 
atoms acquire orbital degrees of freedom, 
as each lattice site contains localized Wannier orbitals---$s$, $p$, $d$, 
\ldots~\cite{Li_2016}, whose quantum superposition gives multiple Bloch bands. 
Having atoms condense at lattice quasi-momentum $k=0$, the orbital quantum state 
is described by a density matrix 
\be 
\textstyle \rho = \sum_{\nu \nu'} \rho_{\nu \nu'} |\nu \rangle \langle \nu'|, 
\label{eq:rho}
\ee 
where $|\nu = s, p, d, \ldots\rangle$ represents the Bloch  mode at $k=0$ of the $\nu$-orbital band. 
In order to control the orbital state, we apply lattice modulation with a particular frequency $\omega$ to resonantly couple $s$ and $d$ orbitals (see Fig.~\ref{fig:orqbit}), which introduces an additional potential, 
\be 
\Delta V(x, t)  = A \sin (\omega t + \varphi) V_p (x),
\label{eq:potential} 
\ee 
with both of  amplitude $A$, and phase $\varphi$ programmable in our experiment (see Supplementary Material). 
With leakage to other orbitals neglected, the system corresponds to a two-level system, defining our atom-orbital qubit, with $|d\rangle$ and $|s\rangle$ orbital states identified as the qubit basis states, $|0\rangle$ and $|1\rangle$. 
In our experimental lattice setup, the energy gap between $s$- and $d$-bands has a large detuning from the energy gap 
between other orbitals except the gap from the $s$- to $p$-band. Despite the absence of the energy suppression of the $s$ to $p$ transition, this transition is forbidden due to inversion symmetry present during the lattice modulation. 
The transitions from the $s$-orbital  to the undesired orbitals are thus avoided, either by energy suppression or by symmetry.

For qubit initialization, we selectively load the atomic BEC into the lattice using our previously developed shortcut preparation method~\cite{NJP_zhou}, with which we are able to initialize the qubit to an arbitrary state, 
$\left\lvert \psi \right\rangle=\cos \theta\left\lvert 0\right\rangle +\sin \theta e^{i\phi}\left\lvert 1\right\rangle$, 
within $250$ $\mu$s.   
For qubit readout, we implement a  time-of-flight  quantum state tomography (TOFQST), which extracts the full information of the density matrix $\rho_{\nu \nu'}$ from the atomic momentum distribution (see Supplemental Material). 
In the experiment, we prepare  a set of six complementary initial states, 
$\mathbb{S} \equiv \{  
\left\lvert 0\right\rangle ,
\left\lvert 1\right\rangle,
\left\lvert +\right\rangle=\frac{1}{\sqrt{2}}(\left\lvert 0\right\rangle +\left\lvert 1\right\rangle),
\left\lvert -\right\rangle=\frac{1}{\sqrt{2}}(\left\lvert 0\right\rangle -\left\lvert 1\right\rangle),
\left\lvert +i\right\rangle=\frac{1}{\sqrt{2}}(\left\lvert 0\right\rangle +i\left\lvert 1\right\rangle),
\left\lvert -i\right\rangle=\frac{1}{\sqrt{2}}(\left\lvert 0\right\rangle -i\left\lvert 1\right\rangle) \}$, 
whose measured state fidelities are $99.96(2)\%$, $99.88(5)\%$, $99.30(11)$, $99.37(10)\%$, $99.83(5)\%$, and $99.98(1)\%$.  The averaged fidelity is $99.72(7)\%$.

{
By initializing to the excited $d$-orbital state, namely $|0\rangle$, we measure the time ($t$) evolution of the atom number $N_d$ in the $d$ band, and obtain $N_d (t)$. Fitting this function to $e^{-t/T_{\rm relax}}$, we extract the relaxation $T_{\rm relax}$ time, which is $4.5\pm0.1$ ms (Fig.~\ref{fig:orqbit}(c)).}
To measure the dephasing time $T_2$, we prepare an orbital superposition state $\left[ |0\rangle +|1\rangle \right]/\sqrt{2}$. For the energy splitting between $s$, and $d$ orbital bands in the lattice, this superposition state develops Ramsey interference fringes in dynamics. The time dependence of the off-diagonal term has a form of  (Fig.~\ref{fig:orqbit}(d))
\be 
{\rm Re} (\rho_{sd} (t))\sim e^{-t/T_2} \cos (\Omega t +\phi),
\ee 
with $\Omega$, $\phi$, $T_2$ determined by fitting the experimental data gotten by band mapping method. We obtain a $T_2$ time  of $2.1\pm0.1$ ms.

\begin{figure}[htp]
\centering
\includegraphics[width=.45\textwidth]{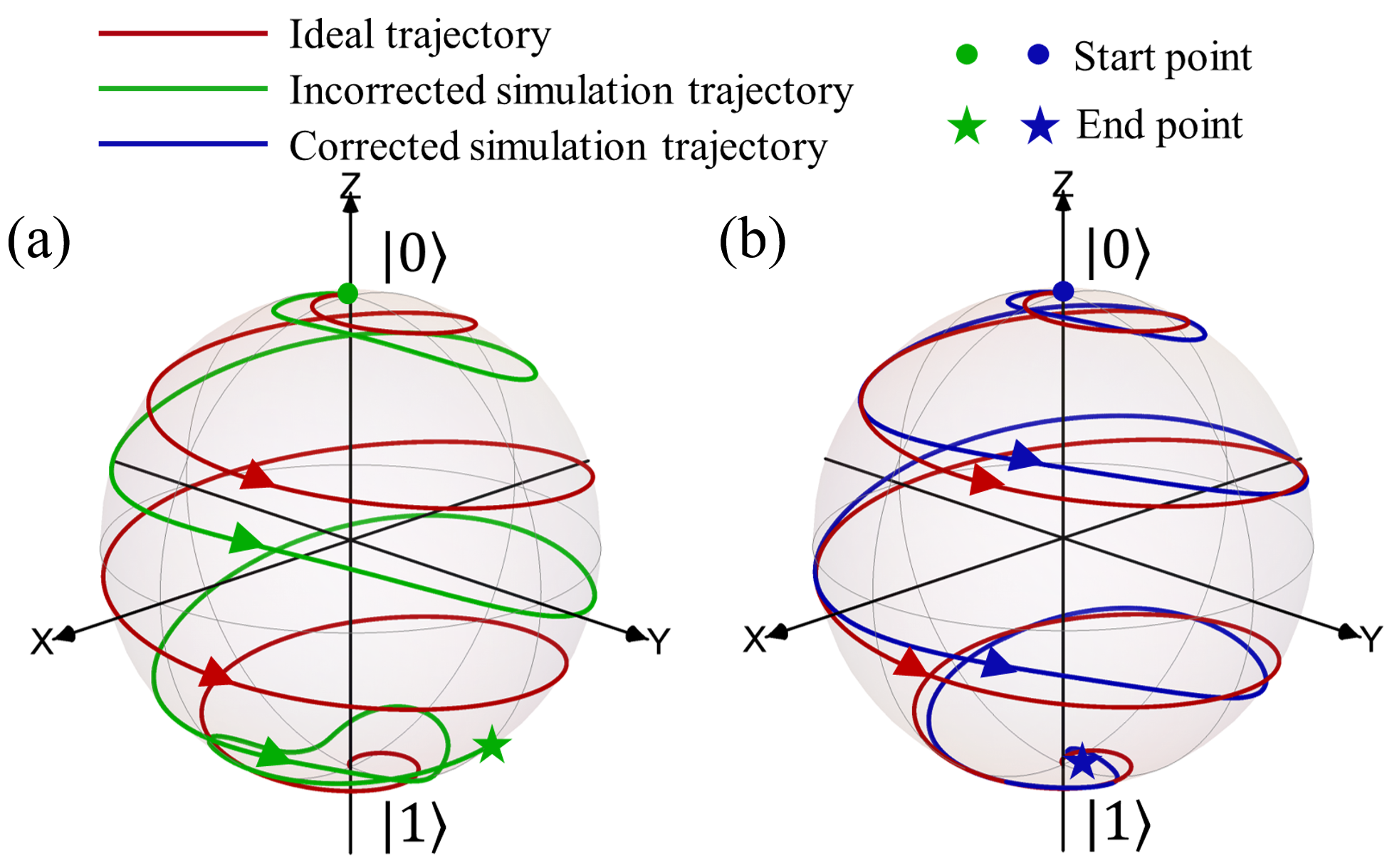}
\caption{{Simulated} time evolution of the $|0\rangle$ state on the Bloch sphere under the holonomic $X$ gate. North and south poles of Bloch sphere denote the $|d\rangle$ and $|s\rangle$ orbital states. The red curve shows the ideal state trajectory under the holonomic control according to the idealized qubit model (Eq.~\eqref{eq:controlHam}). The blue (green) curve shows the state  trajectory by the multi-orbital numerical simulation,
with (without) the orbital leakage error eliminated.}
\label{fig:bloch}
\end{figure}

{\it Nonadiabatic holonomic orbital gate construction.---}
With the optical potential in Eq.~\eqref{eq:potential}, we have a qubit control Hamiltonian $H(t)$ of the form, 
\be 
\label{eq:controlHam}
\begin{aligned}
\frac{1}{2} \Delta \sigma_{z}+\frac{1}{2} \lambda \left(-\cos (\omega t+\varphi) \sigma_{y}+\sin (\omega t+\varphi) \sigma_{x}\right), 
\end{aligned}
\ee 
where $\Delta$ is the gap between the $s$ and $d$-bands at  quasi-momentum $k=0$ ($\Delta = 5.23 E_r$ in our experiment), and the induced coupling by the lattice modulation  is 
\be 
\textstyle \lambda =A\int dx V_{p}(x) \phi_{d}(x)  \phi_{s}(x), 
\ee 
with $\phi_{\nu = s,d}(x)$ the Bloch function of $\nu$-orbital band at $k= 0$. 
With a resonant coupling $\omega \approx \Delta$, the overall quantum gate operation time is determined by $1/\lambda$. To avoid 
{non-resonant} 
transitions to other bands, predominately the $g$-band, which has an energy gap from the $s$-band, $\Delta_{sg}$ 
{($=16.73E_r\gg \Delta$),} 
it is required that $\lambda \ll |\Delta_{sg} - \omega|$, which sets an upper limit for the amplitude of lattice modulation. For this requirement, it becomes more desirable to construct nonadiabatic quantum gates, as the adiabatic control would be too slow---the total adiabatic evolution time is required to be much longer than $1/\lambda$ to maintain quantum adiabaticity. 
 
 We implement  nonadiabatic holonomic orbital control based on 
 a dynamical invariant of the Hamiltonian in Eq.~\eqref{eq:controlHam},  
$I=(\Delta-\omega) \sigma_{z}+\lambda \left(-\cos (\omega t+\varphi) \sigma_{y}+\sin (\omega t+\varphi) \sigma_{x}\right)$~\cite{1969_LewisJMP,2014_Gungordu_JPSJ}, with its instantaneous eigenstates $|\psi_\pm (t) \rangle$. Through one period ($T = 2\pi/\omega$) of quantum evolution, an initial  quantum state  $\sum_n c_n |\psi_n \rangle$ is transformed to $\sum_n c_n e^{i(\gamma_n^g + \gamma_n^d)} |\psi_n \rangle$, with the accumulated geometrical phase $\gamma_n^g = \int_{0}^{T} dt \left\langle\psi_{n}(t)|i\partial_t| \psi_{n}(t)\right\rangle $, and the dynamical phase $\gamma_{n}^{d}=-\int_{0}^{T} dt \left\langle\psi_{n}(t)|H(t)| \psi_{n}(t)\right\rangle $. 
The wavefunction evolution in the dynamical invariant eigenbasis during the holonomic quantum control resembles quantum adiabatic dynamics~\cite{1969_LewisJMP,2014_Gungordu_JPSJ}.
The geometrical phase only depends on the solid angle enclosed by the evolution path of $|\psi_n(t)\rangle$ on the Bloch sphere~\cite{1984_BerryPhase}. 
The holonomic gate is thus a generalization of adiabatic geometrical gate~\cite{2002_Zhu_PRL}. 
This gate has intrinsic resilience against experimental control errors for geometrical protection~\cite{2012_Johansson_PRA,2013_Berger_PRA}, which has been demonstrated with liquid NMR~\cite{2013_Long_PRL,long_2017}, solid-state~\cite{2014_Duan_Nature,2014_Balasubramanian_NC,2016_Yale_NatPhoto,2017_Hideo_Nat_photo,npj_2018,duan_prl_2019}, neutral atoms~\cite{Zhao_PRA_atom_2017,OL_atom_2019,Liu_PRR_atom_2020}, ions~\cite{Duan_science,PhysRevA.87.052307,leibfried_experimental_2003,PhysRevApplied.14.054062,PhysRevLett.127.030502} and superconducting qubits~\cite{2013_Filipp_Nature,2019_Yu_PRL,2019_Yu_PRL,PhysRevLett.121.110501,PhysRevLett.124.230503,Zhang_2019,Zhao_2021}. 
To  exploit the geometrical protection, the dynamical phase has to be cancelled, which corresponds to 
\begin{equation}
\textstyle \lambda^{2}+\Delta(\Delta-\omega)=0.
\label{eq:holonomic_con}
\end{equation}
This gives  the holonomic gate condition~\cite{2014_Gungordu_JPSJ},  which is explicitly satisfied in our lattice modulation design. 
The SU(2) rotation through one period of lattice modulation is then 
$U_{\beta\varphi}(T) = 
\sum_{\pm} e^{i \gamma_{\pm}^g } 
|\psi_{\pm}(0) \rangle \langle\psi_{\pm}(0)|$. This single-qubit rotation can be rewritten as 
\be 
\textstyle U_{\beta\varphi} =-e^{i \pi \sin \beta\left[-\sin \varphi \cos \beta \sigma_{x}+\cos \varphi \cos \beta \sigma_{y}+\sin \beta \sigma_{z}\right]}
\ee 
with $\beta$ determined by the lattice modulation frequency---$\cos ^{2} \beta=\Delta / \omega$ ($\beta \in[0, \pi / 2]$), and the angle $\varphi$ controllable by programming  the potential (Eq.~\eqref{eq:potential}). 
An arbitrary target of holonomic SU(2) rotation, $U_{\rm target}$, is approached by  combining multiple periods ($M$ times) of lattice modulation, which gives a concatenated unitary 
$
U=U_{\beta_M \varphi_{M}}U_{\beta_{M-1} \varphi_{M-1}}...U_{\beta_2\varphi_{2} }U_{\beta_1\varphi_{1}}$. 
The parameter sequence ($\beta_j, \varphi_j$) is determined by optimizing the gate fidelity
$
F=| \operatorname{tr}\left(U^{\dagger} U_{\text { target }}\right) | /2
$. 
We then obtain a  control sequence of lattice modulation frequency, amplitude,  and phase, denoted by $\Theta_j \equiv (\omega_j, A_j,  \varphi_j)$. 
With this scheme, we {expect gate fidelities} above $99.999\%$ using $M \le 5$ for $X$, $Y$,  $Z$, Hadamard, and $\pi$/8 gates under the idealized model in Eq.~\eqref{eq:controlHam}).

 \begin{figure}[htp]
\centering
\includegraphics[width=.48\textwidth]{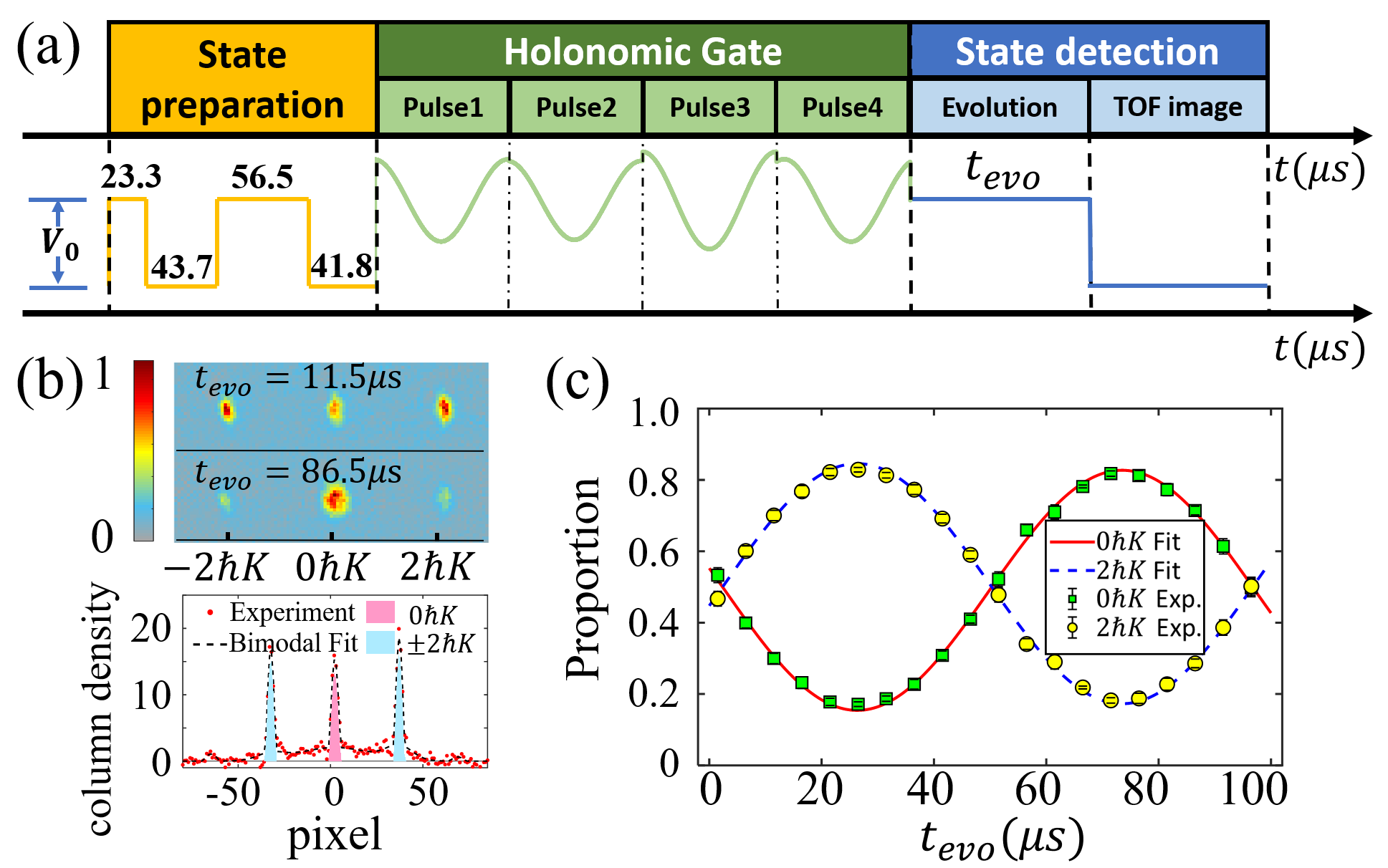}
\caption{Experimental realization of the holonomic quantum gates.  
(a) Experimental procedures
of performing holonomic $X$ gate on the initial state, $\left\lvert -i\right\rangle $. 
The numbers on top of the square pulses in (a) represent the time durations in units of microseconds. 
(b) Time-of-flight patterns at different evolution time. 
The condensed numbers of atoms at momentum $0\hbar K$, and 
$\pm 2\hbar K$ are extracted following standard analysis in cold atom experiments. 
(c) Oscillation dynamics of the atom number proportion 
at $0\hbar K$ and $\pm 2\hbar K$. We average over four 
experimental runs for each data point, with the error 
bar denoting the standard statistical error. Here we set 
the lattice confinement $V_0 = 5E_r$. 
}
\label{fig:Xgatedemo}
\end{figure}

\begin{table*}[htp]
\begin{tabular}{cccccccc}
\hline
{Initial state}      & {$|0\rangle$}    & {$|1\rangle$}    & {$|+\rangle$}    & {$|-\rangle$}    & {$|+i\rangle$}    & {$|-i\rangle$}   & {Average}\\\hline
{X-gate}             & {99.92(3)}      & {99.08(8)}       & {98.87(11)}                & {93.28(34)}                & {98.75(9)}                 & {99.13(7)}                  & {98.17(16)}\\
{Y-gate}             & {99.19(13)}      & {99.71(6)}       & {98.29(13)}                & {96.91(20)}                & {98.73(11)}                 & {99.15(8)}                  & {98.66(13)}\\
{Z-gate}             & {97.87(14)}      & {99.97(3)}       & {97.47(18)}                & {94.83(29)}                & {99.14(11)}                 & {97.75(16)}                 & {97.84(17)}\\
{Hadamard}           & {98.24(18)}      & {98.38(11)}      & {99.45(7)}                 & {97.90(20)}                & {99.08(8)}                  & {96.77(23)}                 & {98.30(16)}\\
{$\pi$/8-gate}           & {99.21(21)}      & {99.76(7)}      & {97.78(30)}                 & {95.94(41)}                & {99.69(8)}                  & {99.28(24)}                 & {98.61(25)}\\
\hline
\end{tabular}
\caption{Measured fidelities of the experimental holonomic quantum gates  by TOFQST.}
\label{tab:gatefidelity}
\end{table*}

 However, in comparing the results of the ideal model with a more precise 
 multi-orbital numerical simulation that incorporates all continuous degrees of 
 freedom of the lattice (Supplementary Material), we find significant 
 discrepancy. Fig.~\ref{fig:bloch} illustrates one example 
 of state evolution under a constructed holonomic $X$ gate.  
 In the example shown in Fig.~\ref{fig:bloch}(a), the 
 gate fidelity obtained by the multi-orbital numerical 
 simulation is below $85\%$. The sizeable difference from the ideal model 
 is attributed to the leakage to other unwanted orbitals.
 Even a small fraction of $g$-band population below $5\%$ is found to strongly 
 disturb  the qubit evolution on the Bloch sphere (Fig.~\ref{fig:bloch}(a)). 
 To resolve the problem of  orbital leakage, we develop an orbital leakage elimination protocol ({\it Supplemental Material}), where the leakage error is minimized by optimizing the control sequence $\Theta_j$.  
 With the optimal control sequence  designed for holonomic $X$, $Y$, $Z$, Hadamard, and $\pi/8$ gates ({\it Supplemental Material}), the gate fidelity is improved to above $98\%$  in the multi-orbital numerical simulation. The orbital leakage elimination protocol is implemented in our experiment for high fidelity realization of holonomic gates. 
 

\begin{figure}[htbp]
\centering
\includegraphics[width=.48\textwidth]{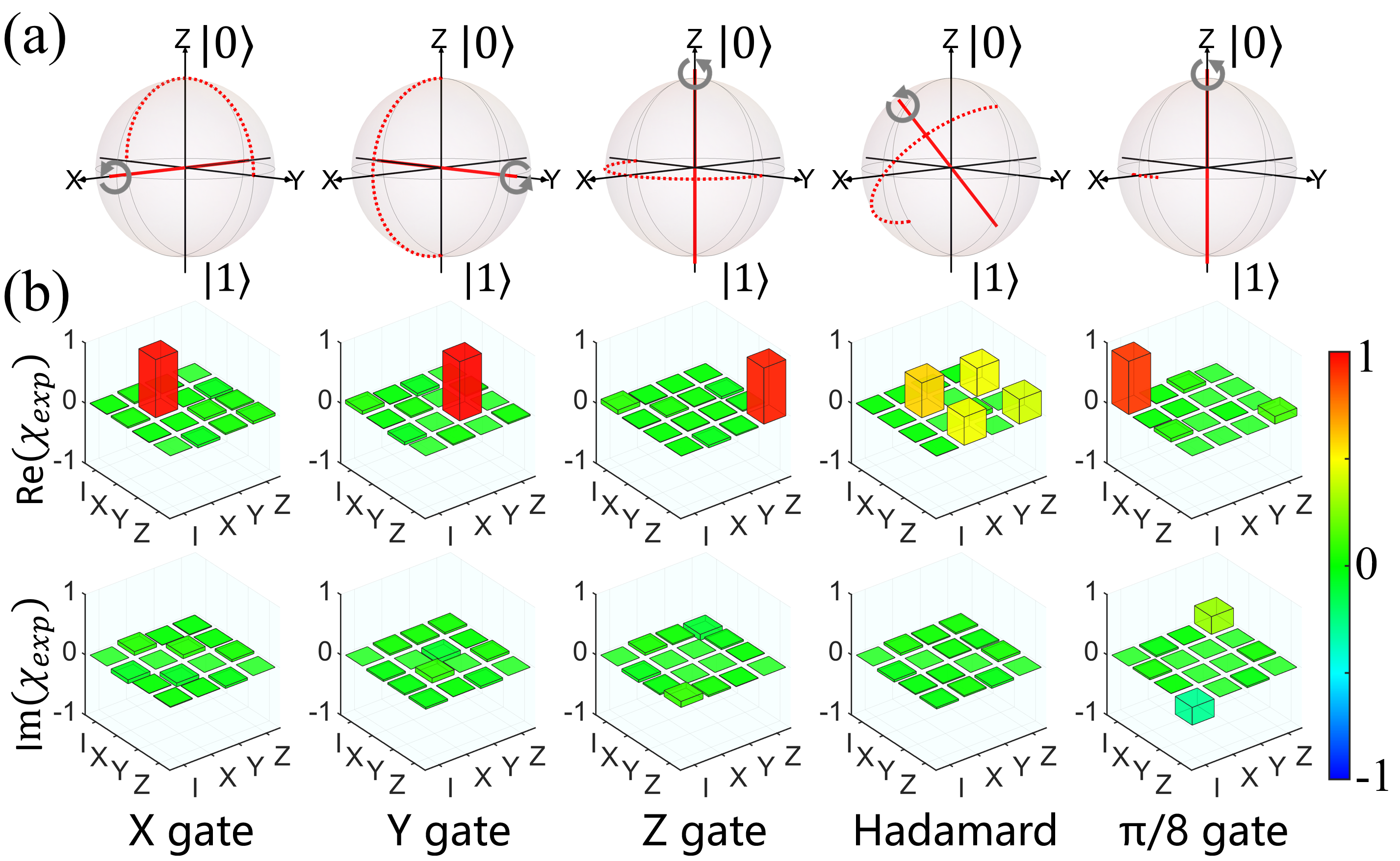}
\caption{Representations of the experimentally accomplished  quantum processes. (a) The
schematic illustration of the quantum processes. 
The red lines, gray arrows, and red dots, represent the rotation axis,
direction and angle of the corresponding SU(2) rotation, respectively. 
(b) Process matrices of the implemented holonomic $X$,
$Y$, $Z$, Hadamard, and $\pi/8$ gates by QPT measurements. 
}
\label{fig5}
\end{figure}

{\it Holonomic quantum orbital gate fidelities.---}  
In the experiment, we implement holonomic orbital gates taking the control sequences as designed with our orbital leakage elimination protocol. The band population dynamics during the gate operation has a very good agreement with our multi-orbital numerical simulation ({\it Supplementary Figure~S5}). 
 Fig.~\ref{fig:Xgatedemo} shows the experimental realization of the holonomic $X$ gate. The atomic BEC is initialized in the quantum state $|-i\rangle$. We then perform the holonomic $X$ gate control. The occupation of momentum $\tilde{p}=0$ and $\tilde{p} = 2\hbar K$ (with $K = 2\pi/l$) states are measured through time-of-flight (TOF), by which the full orbital qubit state is mapped out through TOFQST ({\it Supplemental Material}).  
This experimental procedure is repeated for all the initial states in $\mathbb{S}$. The fidelity ($F$) averaged over the final states is  $98.17(16)\%$ (Table~\ref{tab:gatefidelity}), 
as calculated from the overlap of density matrix measured in the experiment ($\rho_{\rm exp}$) 
with the theoretical construction ($\rho_{\rm th}$), i.e., $F = \left\lvert {\rm Tr} (\rho_{\rm exp}\rho_{th}^{\dagger})\right\rvert/\sqrt{{\rm Tr}(\rho_{\rm exp}\rho_{\rm exp}^{\dagger}){\rm Tr}(\rho_{\rm th}\rho_{\rm th}^{\dagger})} $. 
We also implement the holonomic  $Y$, $Z$, Hadamard, and $\pi/8$ gates in the experiment, whose averaged final state fidelities are  $98.66(13)\%$, $97.84(17)\%$, $98.30(16)\%$, and $98.61(25)\%$. 


One way to construct universal holonomic gate for single qubit rotation is to combine Hadamard and $\pi/8$ gates~\cite{1999_Boykin_universal}. The other is to search for the control sequence numerically, whose  universality is confirmed through our multi-orbital numerical simulation. We randomly sample a hundred Haar random SU(2) rotations, and find all can be constructed by the concatenated unitary sequence with $M\le 5$, producing final state  fidelities all above $98\%$ ({\it Supplementary Figure~S2}).

To characterize the holonomic quantum gates  directly, we further perform quantum process tomography (QPT).  
We initialize the atomic BEC in the six orbital states in $\mathbb{S}$, 
and map out the final quantum states following the holonomic quantum gates by TOFQST. 
The $\chi$ matrices, which represent the quantum gate operations, are reconstructed  by the QPT method~\cite{nielsen_chuang_2010,Howard_2006}. 
The results are shown in Fig.~\ref{fig5}. For completeness we also provide   quantum process fidelities (QPFs) as defined by  
$\left\lvert {\rm Tr}(\chi_{\rm exp}\chi_{\rm th}^{\dagger})\right\rvert/\sqrt{{\rm Tr}(\chi_{\rm exp}\chi_{exp}^{\dagger}){\rm Tr}(\chi_{\rm th}\chi_{\rm th}^{\dagger})} $, with $\chi_{\rm exp}$ and $\chi_{\rm th}$, 
the $\chi$ matrix reconstructed from experimental measurements and the theoretical expectation, respectively. 
The measured  QPFs  are $98.47(9)\%$, $98.35(11)\%$, $97.81(13)\%$, $98.53(8)\%$, $98.63(15)\%$, for the X, Y, Z, Hadamard, and $\pi$/8 gates in the experiment. 


{\it Discussion.---} 
We have constructed an orbital qubit using atomic Bose-Einstein condensation in a one-dimensional optical lattice. The atom orbital qubit is equipped with nonadiabatic holonomic quantum control, that shows noise resilience against laser intensity fluctuations due to geometrical protection ({\it Supplementary Figure~S3}). In order to further improve the gate fidelity in the experiment, we expect it is helpful to incorporate  the inhomogenity  produced by the trap, which can be treated by integrating spin echo like pulse or optimal control protocols~\cite{2013_Urabe_Holonomic,Xuezhengyuan_2019,2020_Ahn_PRR}. 

To achieve universal quantum computing with the orbit qubit setup, we still need to construct two-qubit gates, which can be implemented through an atom collision scheme ({\it Supplementary Material}). We expect the orbital quantum control techniques developed here can be generalized to full quantum control of all orbitals. For orbital transitions that respect parity symmetry, one may still take the lattice modulation protocol but using different resonant frequencies. To control orbital transitions violating parity symmetry would require lattice shaking techniques~\cite{2013_Chin_Shaking}. 
The orbital lattice based quantum computing would provide novel opportunities for  programmable quantum simulations. For example, the demonstrated arbitrary coupling between $s$ and $d$ orbitals may be used to perform programmable  quantum simulations of spin-orbital interaction analogue and the consequent topological physics~\cite{2013_Li_NC}.  
 

{\it Acknowledgements.---}
We acknowledge helpful discussion with Yidun Wan. 
This work is supported by National Natural Science Foundation of China (Grants No. 61727819, 11934002, 91736208, and 11774067), National Program on Key Basic Research Project of China (Grant No. 2016YFA0301501, Grant No. 2017YFA0304204), Shanghai Municipal Science and Technology Major Project (Grant No. 2019SHZDZX01). 

$^{\blue \ddagger}$ These authors contributed equally to this work.

\bibliography{do}

\begin{thebibliography}{67}%
\makeatletter
\providecommand \@ifxundefined [1]{%
 \@ifx{#1\undefined}
}%
\providecommand \@ifnum [1]{%
 \ifnum #1\expandafter \@firstoftwo
 \else \expandafter \@secondoftwo
 \fi
}%
\providecommand \@ifx [1]{%
 \ifx #1\expandafter \@firstoftwo
 \else \expandafter \@secondoftwo
 \fi
}%
\providecommand \natexlab [1]{#1}%
\providecommand \enquote  [1]{``#1''}%
\providecommand \bibnamefont  [1]{#1}%
\providecommand \bibfnamefont [1]{#1}%
\providecommand \citenamefont [1]{#1}%
\providecommand \href@noop [0]{\@secondoftwo}%
\providecommand \href [0]{\begingroup \@sanitize@url \@href}%
\providecommand \@href[1]{\@@startlink{#1}\@@href}%
\providecommand \@@href[1]{\endgroup#1\@@endlink}%
\providecommand \@sanitize@url [0]{\catcode `\\12\catcode `\$12\catcode
  `\&12\catcode `\#12\catcode `\^12\catcode `\_12\catcode `\%12\relax}%
\providecommand \@@startlink[1]{}%
\providecommand \@@endlink[0]{}%
\providecommand \url  [0]{\begingroup\@sanitize@url \@url }%
\providecommand \@url [1]{\endgroup\@href {#1}{\urlprefix }}%
\providecommand \urlprefix  [0]{URL }%
\providecommand \Eprint [0]{\href }%
\providecommand \doibase [0]{http://dx.doi.org/}%
\providecommand \selectlanguage [0]{\@gobble}%
\providecommand \bibinfo  [0]{\@secondoftwo}%
\providecommand \bibfield  [0]{\@secondoftwo}%
\providecommand \translation [1]{[#1]}%
\providecommand \BibitemOpen [0]{}%
\providecommand \bibitemStop [0]{}%
\providecommand \bibitemNoStop [0]{.\EOS\space}%
\providecommand \EOS [0]{\spacefactor3000\relax}%
\providecommand \BibitemShut  [1]{\csname bibitem#1\endcsname}%
\let\auto@bib@innerbib\@empty
\bibitem [{\citenamefont {Bartlett}\ and\ \citenamefont
  {Musiał}(2007)}]{2007_Bartlett_RMP}%
  \BibitemOpen
  \bibfield  {author} {\bibinfo {author} {\bibfnamefont {R.}~\bibnamefont
  {Bartlett}}\ and\ \bibinfo {author} {\bibfnamefont {M.}~\bibnamefont
  {Musiał}},\ }\href {\doibase 10.1103/RevModPhys.79.291} {\bibfield
  {journal} {\bibinfo  {journal} {Rev. Mod. Phys.}\ }\textbf {\bibinfo {volume}
  {79}},\ \bibinfo {pages} {291} (\bibinfo {year} {2007})}\BibitemShut
  {NoStop}%
\bibitem [{\citenamefont {Hwang}\ \emph {et~al.}(2012)\citenamefont {Hwang},
  \citenamefont {Iwasa}, \citenamefont {Kawasaki}, \citenamefont {Keimer},
  \citenamefont {Nagaosa},\ and\ \citenamefont {Tokura}}]{2012_Hwang_NatMat}%
  \BibitemOpen
  \bibfield  {author} {\bibinfo {author} {\bibfnamefont {H.}~\bibnamefont
  {Hwang}}, \bibinfo {author} {\bibfnamefont {Y.}~\bibnamefont {Iwasa}},
  \bibinfo {author} {\bibfnamefont {M.}~\bibnamefont {Kawasaki}}, \bibinfo
  {author} {\bibfnamefont {B.}~\bibnamefont {Keimer}}, \bibinfo {author}
  {\bibfnamefont {N.}~\bibnamefont {Nagaosa}}, \ and\ \bibinfo {author}
  {\bibfnamefont {Y.}~\bibnamefont {Tokura}},\ }\href
  {https://doi.org/10.1038/nmat3223} {\bibfield  {journal} {\bibinfo  {journal}
  {Nat. Mater.}\ }\textbf {\bibinfo {volume} {11}},\ \bibinfo {pages} {103}
  (\bibinfo {year} {2012})}\BibitemShut {NoStop}%
\bibitem [{\citenamefont {Stewart}(2011)}]{2011_Stewart_RMP}%
  \BibitemOpen
  \bibfield  {author} {\bibinfo {author} {\bibfnamefont {G.~R.}\ \bibnamefont
  {Stewart}},\ }\href {\doibase 10.1103/RevModPhys.83.1589} {\bibfield
  {journal} {\bibinfo  {journal} {Rev. Mod. Phys.}\ }\textbf {\bibinfo {volume}
  {83}},\ \bibinfo {pages} {1589} (\bibinfo {year} {2011})}\BibitemShut
  {NoStop}%
\bibitem [{\citenamefont {Li}\ \emph {et~al.}(2020)\citenamefont {Li},
  \citenamefont {Nan},\ and\ \citenamefont {Pan}}]{2020_Li_Chiral}%
  \BibitemOpen
  \bibfield  {author} {\bibinfo {author} {\bibfnamefont {X.}~\bibnamefont
  {Li}}, \bibinfo {author} {\bibfnamefont {J.}~\bibnamefont {Nan}}, \ and\
  \bibinfo {author} {\bibfnamefont {X.}~\bibnamefont {Pan}},\ }\href {\doibase
  10.1103/PhysRevLett.125.263002} {\bibfield  {journal} {\bibinfo  {journal}
  {Phys. Rev. Lett.}\ }\textbf {\bibinfo {volume} {125}},\ \bibinfo {pages}
  {263002} (\bibinfo {year} {2020})}\BibitemShut {NoStop}%
\bibitem [{\citenamefont {Liu}\ \emph {et~al.}(2021)\citenamefont {Liu},
  \citenamefont {Xiao}, \citenamefont {Koo},\ and\ \citenamefont
  {Yan}}]{2021_Yan_NatMat}%
  \BibitemOpen
  \bibfield  {author} {\bibinfo {author} {\bibfnamefont {Y.}~\bibnamefont
  {Liu}}, \bibinfo {author} {\bibfnamefont {J.}~\bibnamefont {Xiao}}, \bibinfo
  {author} {\bibfnamefont {J.}~\bibnamefont {Koo}}, \ and\ \bibinfo {author}
  {\bibfnamefont {B.}~\bibnamefont {Yan}},\ }\href {\doibase
  10.1038/s41563-021-00924-5} {\bibfield  {journal} {\bibinfo  {journal} {Nat.
  Mater.}\ }\textbf {\bibinfo {volume} {20}},\ \bibinfo {pages} {638} (\bibinfo
  {year} {2021})}\BibitemShut {NoStop}%
\bibitem [{\citenamefont {Aspuru-Guzik}\ \emph {et~al.}(2005)\citenamefont
  {Aspuru-Guzik}, \citenamefont {Dutoi}, \citenamefont {Love},\ and\
  \citenamefont {Head-Gordon}}]{aspuru2005simulated}%
  \BibitemOpen
  \bibfield  {author} {\bibinfo {author} {\bibfnamefont {A.}~\bibnamefont
  {Aspuru-Guzik}}, \bibinfo {author} {\bibfnamefont {A.}~\bibnamefont {Dutoi}},
  \bibinfo {author} {\bibfnamefont {P.}~\bibnamefont {Love}}, \ and\ \bibinfo
  {author} {\bibfnamefont {M.}~\bibnamefont {Head-Gordon}},\ }\href
  {https://doi.org/10.1126/science.1113479} {\bibfield  {journal} {\bibinfo
  {journal} {Science}\ }\textbf {\bibinfo {volume} {309}},\ \bibinfo {pages}
  {1704} (\bibinfo {year} {2005})}\BibitemShut {NoStop}%
\bibitem [{\citenamefont {McArdle}\ \emph {et~al.}(2020)\citenamefont
  {McArdle}, \citenamefont {Endo}, \citenamefont {Aspuru-Guzik}, \citenamefont
  {Benjamin},\ and\ \citenamefont {Yuan}}]{2020_McArdle_RMP}%
  \BibitemOpen
  \bibfield  {author} {\bibinfo {author} {\bibfnamefont {S.}~\bibnamefont
  {McArdle}}, \bibinfo {author} {\bibfnamefont {S.}~\bibnamefont {Endo}},
  \bibinfo {author} {\bibfnamefont {A.}~\bibnamefont {Aspuru-Guzik}}, \bibinfo
  {author} {\bibfnamefont {S.}~\bibnamefont {Benjamin}}, \ and\ \bibinfo
  {author} {\bibfnamefont {X.}~\bibnamefont {Yuan}},\ }\href {\doibase
  10.1103/RevModPhys.92.015003} {\bibfield  {journal} {\bibinfo  {journal}
  {Rev. Mod. Phys.}\ }\textbf {\bibinfo {volume} {92}},\ \bibinfo {pages}
  {015003} (\bibinfo {year} {2020})}\BibitemShut {NoStop}%
\bibitem [{\citenamefont {Kandala}\ \emph {et~al.}(2017)\citenamefont
  {Kandala}, \citenamefont {Mezzacapo}, \citenamefont {Temme}, \citenamefont
  {Takita}, \citenamefont {Brink}, \citenamefont {Chow},\ and\ \citenamefont
  {Gambetta}}]{kandala2017hardware}%
  \BibitemOpen
  \bibfield  {author} {\bibinfo {author} {\bibfnamefont {A.}~\bibnamefont
  {Kandala}}, \bibinfo {author} {\bibfnamefont {A.}~\bibnamefont {Mezzacapo}},
  \bibinfo {author} {\bibfnamefont {K.}~\bibnamefont {Temme}}, \bibinfo
  {author} {\bibfnamefont {M.}~\bibnamefont {Takita}}, \bibinfo {author}
  {\bibfnamefont {M.}~\bibnamefont {Brink}}, \bibinfo {author} {\bibfnamefont
  {J.}~\bibnamefont {Chow}}, \ and\ \bibinfo {author} {\bibfnamefont
  {J.}~\bibnamefont {Gambetta}},\ }\href {https://doi.org/10.1038/nature23879}
  {\bibfield  {journal} {\bibinfo  {journal} {Nature (London)}\ }\textbf
  {\bibinfo {volume} {549}},\ \bibinfo {pages} {242} (\bibinfo {year}
  {2017})}\BibitemShut {NoStop}%
\bibitem [{\citenamefont {Quantum}\ \emph {et~al.}(2020)\citenamefont {Quantum}
  \emph {et~al.}}]{google2020hartree}%
  \BibitemOpen
  \bibfield  {author} {\bibinfo {author} {\bibfnamefont {G.~A.}\ \bibnamefont
  {Quantum}} \emph {et~al.},\ }\href {https://doi.org/10.1126/science.abb9811}
  {\bibfield  {journal} {\bibinfo  {journal} {Science}\ }\textbf {\bibinfo
  {volume} {369}},\ \bibinfo {pages} {1084} (\bibinfo {year}
  {2020})}\BibitemShut {NoStop}%
\bibitem [{\citenamefont {Hempel}\ \emph {et~al.}(2018)\citenamefont {Hempel},
  \citenamefont {Maier}, \citenamefont {Romero}, \citenamefont {McClean},
  \citenamefont {Monz}, \citenamefont {Shen}, \citenamefont {Jurcevic},
  \citenamefont {Lanyon}, \citenamefont {Love}, \citenamefont {Babbush},
  \citenamefont {Aspuru-Guzik}, \citenamefont {Blatt},\ and\ \citenamefont
  {Roos}}]{2018_Blatt_PRX}%
  \BibitemOpen
  \bibfield  {author} {\bibinfo {author} {\bibfnamefont {C.}~\bibnamefont
  {Hempel}}, \bibinfo {author} {\bibfnamefont {C.}~\bibnamefont {Maier}},
  \bibinfo {author} {\bibfnamefont {J.}~\bibnamefont {Romero}}, \bibinfo
  {author} {\bibfnamefont {J.}~\bibnamefont {McClean}}, \bibinfo {author}
  {\bibfnamefont {T.}~\bibnamefont {Monz}}, \bibinfo {author} {\bibfnamefont
  {H.}~\bibnamefont {Shen}}, \bibinfo {author} {\bibfnamefont {P.}~\bibnamefont
  {Jurcevic}}, \bibinfo {author} {\bibfnamefont {B.}~\bibnamefont {Lanyon}},
  \bibinfo {author} {\bibfnamefont {P.}~\bibnamefont {Love}}, \bibinfo {author}
  {\bibfnamefont {R.}~\bibnamefont {Babbush}}, \bibinfo {author} {\bibfnamefont
  {A.}~\bibnamefont {Aspuru-Guzik}}, \bibinfo {author} {\bibfnamefont
  {R.}~\bibnamefont {Blatt}}, \ and\ \bibinfo {author} {\bibfnamefont
  {C.}~\bibnamefont {Roos}},\ }\href {\doibase 10.1103/PhysRevX.8.031022}
  {\bibfield  {journal} {\bibinfo  {journal} {Phys. Rev. X}\ }\textbf {\bibinfo
  {volume} {8}},\ \bibinfo {pages} {031022} (\bibinfo {year}
  {2018})}\BibitemShut {NoStop}%
\bibitem [{\citenamefont {Nam}\ \emph {et~al.}(2020)\citenamefont {Nam},
  \citenamefont {Chen}, \citenamefont {Pisenti}, \citenamefont {Wright},
  \citenamefont {Delaney}, \citenamefont {Maslov}, \citenamefont {Brown},
  \citenamefont {Allen}, \citenamefont {Amini}, \citenamefont {Apisdorf} \emph
  {et~al.}}]{nam2020ground}%
  \BibitemOpen
  \bibfield  {author} {\bibinfo {author} {\bibfnamefont {Y.}~\bibnamefont
  {Nam}}, \bibinfo {author} {\bibfnamefont {J.-S.}\ \bibnamefont {Chen}},
  \bibinfo {author} {\bibfnamefont {N.}~\bibnamefont {Pisenti}}, \bibinfo
  {author} {\bibfnamefont {K.}~\bibnamefont {Wright}}, \bibinfo {author}
  {\bibfnamefont {C.}~\bibnamefont {Delaney}}, \bibinfo {author} {\bibfnamefont
  {D.}~\bibnamefont {Maslov}}, \bibinfo {author} {\bibfnamefont
  {K.}~\bibnamefont {Brown}}, \bibinfo {author} {\bibfnamefont
  {S.}~\bibnamefont {Allen}}, \bibinfo {author} {\bibfnamefont
  {J.}~\bibnamefont {Amini}}, \bibinfo {author} {\bibfnamefont
  {J.}~\bibnamefont {Apisdorf}},  \emph {et~al.},\ }\href
  {https://doi.org/10.1038/s41534-020-0259-3} {\bibfield  {journal} {\bibinfo
  {journal} {npj Quantum Inform.}\ }\textbf {\bibinfo {volume} {6}},\ \bibinfo
  {pages} {1} (\bibinfo {year} {2020})}\BibitemShut {NoStop}%
\bibitem [{\citenamefont {McClean}\ \emph {et~al.}(2016)\citenamefont
  {McClean}, \citenamefont {Romero}, \citenamefont {Babbush},\ and\
  \citenamefont {Aspuru-Guzik}}]{2016_McClean_NJP}%
  \BibitemOpen
  \bibfield  {author} {\bibinfo {author} {\bibfnamefont {J.}~\bibnamefont
  {McClean}}, \bibinfo {author} {\bibfnamefont {J.}~\bibnamefont {Romero}},
  \bibinfo {author} {\bibfnamefont {R.}~\bibnamefont {Babbush}}, \ and\
  \bibinfo {author} {\bibfnamefont {A.}~\bibnamefont {Aspuru-Guzik}},\ }\href
  {\doibase 10.1088/1367-2630/18/2/023023} {\bibfield  {journal} {\bibinfo
  {journal} {New J. Phys.}\ }\textbf {\bibinfo {volume} {18}},\ \bibinfo
  {pages} {023023} (\bibinfo {year} {2016})}\BibitemShut {NoStop}%
\bibitem [{\citenamefont {Li}\ and\ \citenamefont {Liu}(2016)}]{Li_2016}%
  \BibitemOpen
  \bibfield  {author} {\bibinfo {author} {\bibfnamefont {X.}~\bibnamefont
  {Li}}\ and\ \bibinfo {author} {\bibfnamefont {W.}~\bibnamefont {Liu}},\
  }\href {\doibase 10.1088/0034-4885/79/11/116401} {\bibfield  {journal}
  {\bibinfo  {journal} {Rep. Prog. Phys.}\ }\textbf {\bibinfo {volume} {79}},\
  \bibinfo {pages} {116401} (\bibinfo {year} {2016})}\BibitemShut {NoStop}%
\bibitem [{\citenamefont {Isacsson}\ and\ \citenamefont
  {Girvin}(2005)}]{2005_Girvin_PRA}%
  \BibitemOpen
  \bibfield  {author} {\bibinfo {author} {\bibfnamefont {A.}~\bibnamefont
  {Isacsson}}\ and\ \bibinfo {author} {\bibfnamefont {S.}~\bibnamefont
  {Girvin}},\ }\href {\doibase 10.1103/PhysRevA.72.053604} {\bibfield
  {journal} {\bibinfo  {journal} {Phys. Rev. A}\ }\textbf {\bibinfo {volume}
  {72}},\ \bibinfo {pages} {053604} (\bibinfo {year} {2005})}\BibitemShut
  {NoStop}%
\bibitem [{\citenamefont {Liu}\ and\ \citenamefont {Wu}(2006)}]{2006_Liu_PRA}%
  \BibitemOpen
  \bibfield  {author} {\bibinfo {author} {\bibfnamefont {W.}~\bibnamefont
  {Liu}}\ and\ \bibinfo {author} {\bibfnamefont {C.}~\bibnamefont {Wu}},\
  }\href {\doibase 10.1103/PhysRevA.74.013607} {\bibfield  {journal} {\bibinfo
  {journal} {Phys. Rev. A}\ }\textbf {\bibinfo {volume} {74}},\ \bibinfo
  {pages} {013607} (\bibinfo {year} {2006})}\BibitemShut {NoStop}%
\bibitem [{\citenamefont {Kuklov}(2006)}]{2006_Kuklov_PRL}%
  \BibitemOpen
  \bibfield  {author} {\bibinfo {author} {\bibfnamefont {A.~B.}\ \bibnamefont
  {Kuklov}},\ }\href {\doibase 10.1103/PhysRevLett.97.110405} {\bibfield
  {journal} {\bibinfo  {journal} {Phys. Rev. Lett.}\ }\textbf {\bibinfo
  {volume} {97}},\ \bibinfo {pages} {110405} (\bibinfo {year}
  {2006})}\BibitemShut {NoStop}%
\bibitem [{\citenamefont {Wirth}\ \emph {et~al.}(2011)\citenamefont {Wirth},
  \citenamefont {Ölschläger},\ and\ \citenamefont
  {Hemmerich}}]{2011_Wirth_NatPhys}%
  \BibitemOpen
  \bibfield  {author} {\bibinfo {author} {\bibfnamefont {G.}~\bibnamefont
  {Wirth}}, \bibinfo {author} {\bibfnamefont {M.}~\bibnamefont {Ölschläger}},
  \ and\ \bibinfo {author} {\bibfnamefont {A.}~\bibnamefont {Hemmerich}},\
  }\href {https://doi.org/10.1038/nphys1857} {\bibfield  {journal} {\bibinfo
  {journal} {Nat. Phys.}\ }\textbf {\bibinfo {volume} {7}},\ \bibinfo {pages}
  {147} (\bibinfo {year} {2011})}\BibitemShut {NoStop}%
\bibitem [{\citenamefont {Soltan-Panahi}\ \emph {et~al.}(2012)\citenamefont
  {Soltan-Panahi}, \citenamefont {Lühmann}, \citenamefont {Struck},
  \citenamefont {Windpassinger},\ and\ \citenamefont
  {Sengstock}}]{2012_Sengstock_NatPhys}%
  \BibitemOpen
  \bibfield  {author} {\bibinfo {author} {\bibfnamefont {P.}~\bibnamefont
  {Soltan-Panahi}}, \bibinfo {author} {\bibfnamefont {D.-S.}\ \bibnamefont
  {Lühmann}}, \bibinfo {author} {\bibfnamefont {J.}~\bibnamefont {Struck}},
  \bibinfo {author} {\bibfnamefont {P.}~\bibnamefont {Windpassinger}}, \ and\
  \bibinfo {author} {\bibfnamefont {K.}~\bibnamefont {Sengstock}},\ }\href
  {https://doi.org/10.1038/nphys2128} {\bibfield  {journal} {\bibinfo
  {journal} {Nat. Phys.}\ }\textbf {\bibinfo {volume} {8}},\ \bibinfo {pages}
  {71} (\bibinfo {year} {2012})}\BibitemShut {NoStop}%
\bibitem [{\citenamefont {Jin}\ \emph {et~al.}(2021)\citenamefont {Jin},
  \citenamefont {Zhang}, \citenamefont {Guo}, \citenamefont {Chen},
  \citenamefont {Zhou},\ and\ \citenamefont {Li}}]{Jin_PRL2021}%
  \BibitemOpen
  \bibfield  {author} {\bibinfo {author} {\bibfnamefont {S.}~\bibnamefont
  {Jin}}, \bibinfo {author} {\bibfnamefont {W.}~\bibnamefont {Zhang}}, \bibinfo
  {author} {\bibfnamefont {X.}~\bibnamefont {Guo}}, \bibinfo {author}
  {\bibfnamefont {X.}~\bibnamefont {Chen}}, \bibinfo {author} {\bibfnamefont
  {X.}~\bibnamefont {Zhou}}, \ and\ \bibinfo {author} {\bibfnamefont
  {X.}~\bibnamefont {Li}},\ }\href {\doibase 10.1103/PhysRevLett.126.035301}
  {\bibfield  {journal} {\bibinfo  {journal} {Phys. Rev. Lett.}\ }\textbf
  {\bibinfo {volume} {126}},\ \bibinfo {pages} {035301} (\bibinfo {year}
  {2021})}\BibitemShut {NoStop}%
\bibitem [{\citenamefont {Sun}\ \emph {et~al.}(2012)\citenamefont {Sun},
  \citenamefont {Liu}, \citenamefont {Hemmerich},\ and\ \citenamefont
  {Das~Sarma}}]{2012_Sun_NatPhys}%
  \BibitemOpen
  \bibfield  {author} {\bibinfo {author} {\bibfnamefont {K.}~\bibnamefont
  {Sun}}, \bibinfo {author} {\bibfnamefont {W.}~\bibnamefont {Liu}}, \bibinfo
  {author} {\bibfnamefont {A.}~\bibnamefont {Hemmerich}}, \ and\ \bibinfo
  {author} {\bibfnamefont {S.}~\bibnamefont {Das~Sarma}},\ }\href
  {https://doi.org/10.1038/nphys2134} {\bibfield  {journal} {\bibinfo
  {journal} {Nat. Phys.}\ }\textbf {\bibinfo {volume} {8}},\ \bibinfo {pages}
  {67} (\bibinfo {year} {2012})}\BibitemShut {NoStop}%
\bibitem [{\citenamefont {Li}\ \emph {et~al.}(2013)\citenamefont {Li},
  \citenamefont {Zhao},\ and\ \citenamefont {Vincent~Liu}}]{2013_Li_NC}%
  \BibitemOpen
  \bibfield  {author} {\bibinfo {author} {\bibfnamefont {X.}~\bibnamefont
  {Li}}, \bibinfo {author} {\bibfnamefont {E.}~\bibnamefont {Zhao}}, \ and\
  \bibinfo {author} {\bibfnamefont {W.}~\bibnamefont {Vincent~Liu}},\ }\href
  {\doibase 10.1038/ncomms2523} {\bibfield  {journal} {\bibinfo  {journal}
  {Nat. Commun.}\ }\textbf {\bibinfo {volume} {4}},\ \bibinfo {pages} {1523}
  (\bibinfo {year} {2013})}\BibitemShut {NoStop}%
\bibitem [{\citenamefont {Liu}\ \emph {et~al.}(2014)\citenamefont {Liu},
  \citenamefont {Li}, \citenamefont {Wu},\ and\ \citenamefont
  {Liu}}]{2014_Liu_NC}%
  \BibitemOpen
  \bibfield  {author} {\bibinfo {author} {\bibfnamefont {B.}~\bibnamefont
  {Liu}}, \bibinfo {author} {\bibfnamefont {X.}~\bibnamefont {Li}}, \bibinfo
  {author} {\bibfnamefont {B.}~\bibnamefont {Wu}}, \ and\ \bibinfo {author}
  {\bibfnamefont {W.}~\bibnamefont {Liu}},\ }\href
  {https://doi.org/10.1038/ncomms6064} {\bibfield  {journal} {\bibinfo
  {journal} {Nat. Commun.}\ }\textbf {\bibinfo {volume} {5}},\ \bibinfo {pages}
  {1} (\bibinfo {year} {2014})}\BibitemShut {NoStop}%
\bibitem [{\citenamefont {Khamehchi}\ \emph {et~al.}(2016)\citenamefont
  {Khamehchi}, \citenamefont {Qu}, \citenamefont {Mossman}, \citenamefont
  {Zhang},\ and\ \citenamefont {Engels}}]{2016_Engels_NC}%
  \BibitemOpen
  \bibfield  {author} {\bibinfo {author} {\bibfnamefont {M.}~\bibnamefont
  {Khamehchi}}, \bibinfo {author} {\bibfnamefont {C.}~\bibnamefont {Qu}},
  \bibinfo {author} {\bibfnamefont {M.}~\bibnamefont {Mossman}}, \bibinfo
  {author} {\bibfnamefont {C.}~\bibnamefont {Zhang}}, \ and\ \bibinfo {author}
  {\bibfnamefont {P.}~\bibnamefont {Engels}},\ }\href {\doibase
  10.1038/ncomms10867} {\bibfield  {journal} {\bibinfo  {journal} {Nat.
  Commun.}\ }\textbf {\bibinfo {volume} {7}},\ \bibinfo {pages} {10867}
  (\bibinfo {year} {2016})}\BibitemShut {NoStop}%
\bibitem [{\citenamefont {Lewis}\ and\ \citenamefont
  {Riesenfeld}(1969)}]{1969_LewisJMP}%
  \BibitemOpen
  \bibfield  {author} {\bibinfo {author} {\bibfnamefont {H.}~\bibnamefont
  {Lewis}}\ and\ \bibinfo {author} {\bibfnamefont {W.}~\bibnamefont
  {Riesenfeld}},\ }\href {https://doi.org/10.1063/1.1664991} {\bibfield
  {journal} {\bibinfo  {journal} {J. Math. Phys.}\ }\textbf {\bibinfo {volume}
  {10}},\ \bibinfo {pages} {1458} (\bibinfo {year} {1969})}\BibitemShut
  {NoStop}%
\bibitem [{\citenamefont {Güngördü}\ \emph {et~al.}(2014)\citenamefont
  {Güngördü}, \citenamefont {Wan},\ and\ \citenamefont
  {Nakahara}}]{2014_Gungordu_JPSJ}%
  \BibitemOpen
  \bibfield  {author} {\bibinfo {author} {\bibfnamefont {U.}~\bibnamefont
  {Güngördü}}, \bibinfo {author} {\bibfnamefont {Y.}~\bibnamefont {Wan}}, \
  and\ \bibinfo {author} {\bibfnamefont {M.}~\bibnamefont {Nakahara}},\ }\href
  {\doibase 10.7566/jpsj.83.034001} {\bibfield  {journal} {\bibinfo  {journal}
  {J. Phys. Soc. Jpn.}\ }\textbf {\bibinfo {volume} {83}},\ \bibinfo {pages}
  {034001} (\bibinfo {year} {2014})}\BibitemShut {NoStop}%
\bibitem [{\citenamefont {Zhou}\ \emph {et~al.}(2018)\citenamefont {Zhou},
  \citenamefont {Jin},\ and\ \citenamefont {Schmiedmayer}}]{NJP_zhou}%
  \BibitemOpen
  \bibfield  {author} {\bibinfo {author} {\bibfnamefont {X.}~\bibnamefont
  {Zhou}}, \bibinfo {author} {\bibfnamefont {S.}~\bibnamefont {Jin}}, \ and\
  \bibinfo {author} {\bibfnamefont {J.}~\bibnamefont {Schmiedmayer}},\ }\href
  {http://stacks.iop.org/1367-2630/20/i=5/a=055005} {\bibfield  {journal}
  {\bibinfo  {journal} {New J. Phys.}\ }\textbf {\bibinfo {volume} {20}},\
  \bibinfo {pages} {055005} (\bibinfo {year} {2018})}\BibitemShut {NoStop}%
\bibitem [{\citenamefont {Berry}(1984)}]{1984_BerryPhase}%
  \BibitemOpen
  \bibfield  {author} {\bibinfo {author} {\bibfnamefont {M.}~\bibnamefont
  {Berry}},\ }\href {https://doi.org/10.1098/rspa.1984.0023} {\bibfield
  {journal} {\bibinfo  {journal} {Proc. R. Soc. A-Math. Phys. Eng. Sci.}\
  }\textbf {\bibinfo {volume} {392}},\ \bibinfo {pages} {45} (\bibinfo {year}
  {1984})}\BibitemShut {NoStop}%
\bibitem [{\citenamefont {Zhu}\ and\ \citenamefont
  {Wang}(2002)}]{2002_Zhu_PRL}%
  \BibitemOpen
  \bibfield  {author} {\bibinfo {author} {\bibfnamefont {S.-L.}\ \bibnamefont
  {Zhu}}\ and\ \bibinfo {author} {\bibfnamefont {Z.}~\bibnamefont {Wang}},\
  }\href {\doibase 10.1103/PhysRevLett.89.097902} {\bibfield  {journal}
  {\bibinfo  {journal} {Phys. Rev. Lett.}\ }\textbf {\bibinfo {volume} {89}},\
  \bibinfo {pages} {097902} (\bibinfo {year} {2002})}\BibitemShut {NoStop}%
\bibitem [{\citenamefont {Johansson}\ \emph {et~al.}(2012)\citenamefont
  {Johansson}, \citenamefont {Sjöqvist}, \citenamefont {Andersson},
  \citenamefont {Ericsson}, \citenamefont {Hessmo}, \citenamefont {Singh},\
  and\ \citenamefont {Tong}}]{2012_Johansson_PRA}%
  \BibitemOpen
  \bibfield  {author} {\bibinfo {author} {\bibfnamefont {M.}~\bibnamefont
  {Johansson}}, \bibinfo {author} {\bibfnamefont {E.}~\bibnamefont
  {Sjöqvist}}, \bibinfo {author} {\bibfnamefont {L.}~\bibnamefont
  {Andersson}}, \bibinfo {author} {\bibfnamefont {M.}~\bibnamefont {Ericsson}},
  \bibinfo {author} {\bibfnamefont {B.}~\bibnamefont {Hessmo}}, \bibinfo
  {author} {\bibfnamefont {K.}~\bibnamefont {Singh}}, \ and\ \bibinfo {author}
  {\bibfnamefont {D.}~\bibnamefont {Tong}},\ }\href
  {https://link.aps.org/doi/10.1103/PhysRevA.86.062322} {\bibfield  {journal}
  {\bibinfo  {journal} {Phys. Rev. A}\ }\textbf {\bibinfo {volume} {86}},\
  \bibinfo {pages} {062322} (\bibinfo {year} {2012})}\BibitemShut {NoStop}%
\bibitem [{\citenamefont {Berger}\ \emph {et~al.}(2013)\citenamefont {Berger},
  \citenamefont {Pechal}, \citenamefont {Abdumalikov}, \citenamefont {Eichler},
  \citenamefont {Steffen}, \citenamefont {Fedorov}, \citenamefont {Wallraff},\
  and\ \citenamefont {Filipp}}]{2013_Berger_PRA}%
  \BibitemOpen
  \bibfield  {author} {\bibinfo {author} {\bibfnamefont {S.}~\bibnamefont
  {Berger}}, \bibinfo {author} {\bibfnamefont {M.}~\bibnamefont {Pechal}},
  \bibinfo {author} {\bibfnamefont {A.}~\bibnamefont {Abdumalikov}}, \bibinfo
  {author} {\bibfnamefont {C.}~\bibnamefont {Eichler}}, \bibinfo {author}
  {\bibfnamefont {L.}~\bibnamefont {Steffen}}, \bibinfo {author} {\bibfnamefont
  {A.}~\bibnamefont {Fedorov}}, \bibinfo {author} {\bibfnamefont
  {A.}~\bibnamefont {Wallraff}}, \ and\ \bibinfo {author} {\bibfnamefont
  {S.}~\bibnamefont {Filipp}},\ }\href
  {https://link.aps.org/doi/10.1103/PhysRevA.87.060303} {\bibfield  {journal}
  {\bibinfo  {journal} {Phys. Rev. A}\ }\textbf {\bibinfo {volume} {87}},\
  \bibinfo {pages} {060303} (\bibinfo {year} {2013})}\BibitemShut {NoStop}%
\bibitem [{\citenamefont {Feng}\ \emph {et~al.}(2013)\citenamefont {Feng},
  \citenamefont {Xu},\ and\ \citenamefont {Long}}]{2013_Long_PRL}%
  \BibitemOpen
  \bibfield  {author} {\bibinfo {author} {\bibfnamefont {G.}~\bibnamefont
  {Feng}}, \bibinfo {author} {\bibfnamefont {G.}~\bibnamefont {Xu}}, \ and\
  \bibinfo {author} {\bibfnamefont {G.}~\bibnamefont {Long}},\ }\href {\doibase
  10.1103/PhysRevLett.110.190501} {\bibfield  {journal} {\bibinfo  {journal}
  {Phys. Rev. Lett.}\ }\textbf {\bibinfo {volume} {110}},\ \bibinfo {pages}
  {190501} (\bibinfo {year} {2013})}\BibitemShut {NoStop}%
\bibitem [{\citenamefont {Li}\ \emph {et~al.}(2017)\citenamefont {Li},
  \citenamefont {Liu},\ and\ \citenamefont {Long}}]{long_2017}%
  \BibitemOpen
  \bibfield  {author} {\bibinfo {author} {\bibfnamefont {H.}~\bibnamefont
  {Li}}, \bibinfo {author} {\bibfnamefont {Y.}~\bibnamefont {Liu}}, \ and\
  \bibinfo {author} {\bibfnamefont {G.}~\bibnamefont {Long}},\ }\href {\doibase
  10.1007/s11433-017-9058-7} {\bibfield  {journal} {\bibinfo  {journal} {Sci.
  China-Phys. Mech. Astron.}\ }\textbf {\bibinfo {volume} {60}},\ \bibinfo
  {pages} {080311} (\bibinfo {year} {2017})}\BibitemShut {NoStop}%
\bibitem [{\citenamefont {Zu}\ \emph {et~al.}(2014)\citenamefont {Zu},
  \citenamefont {Wang}, \citenamefont {He}, \citenamefont {Zhang},
  \citenamefont {Dai}, \citenamefont {Wang},\ and\ \citenamefont
  {Duan}}]{2014_Duan_Nature}%
  \BibitemOpen
  \bibfield  {author} {\bibinfo {author} {\bibfnamefont {C.}~\bibnamefont
  {Zu}}, \bibinfo {author} {\bibfnamefont {W.}~\bibnamefont {Wang}}, \bibinfo
  {author} {\bibfnamefont {L.}~\bibnamefont {He}}, \bibinfo {author}
  {\bibfnamefont {W.}~\bibnamefont {Zhang}}, \bibinfo {author} {\bibfnamefont
  {C.}~\bibnamefont {Dai}}, \bibinfo {author} {\bibfnamefont {F.}~\bibnamefont
  {Wang}}, \ and\ \bibinfo {author} {\bibfnamefont {L.}~\bibnamefont {Duan}},\
  }\href {\doibase 10.1038/nature13729} {\bibfield  {journal} {\bibinfo
  {journal} {Nature (London)}\ }\textbf {\bibinfo {volume} {514}},\ \bibinfo
  {pages} {72} (\bibinfo {year} {2014})}\BibitemShut {NoStop}%
\bibitem [{\citenamefont {Arroyo-Camejo}\ \emph {et~al.}(2014)\citenamefont
  {Arroyo-Camejo}, \citenamefont {Lazariev}, \citenamefont {Hell},\ and\
  \citenamefont {Balasubramanian}}]{2014_Balasubramanian_NC}%
  \BibitemOpen
  \bibfield  {author} {\bibinfo {author} {\bibfnamefont {S.}~\bibnamefont
  {Arroyo-Camejo}}, \bibinfo {author} {\bibfnamefont {A.}~\bibnamefont
  {Lazariev}}, \bibinfo {author} {\bibfnamefont {S.}~\bibnamefont {Hell}}, \
  and\ \bibinfo {author} {\bibfnamefont {G.}~\bibnamefont {Balasubramanian}},\
  }\href {\doibase 10.1038/ncomms5870} {\bibfield  {journal} {\bibinfo
  {journal} {Nat. Commun.}\ }\textbf {\bibinfo {volume} {5}},\ \bibinfo {pages}
  {4870} (\bibinfo {year} {2014})}\BibitemShut {NoStop}%
\bibitem [{\citenamefont {Yale}\ \emph {et~al.}(2016)\citenamefont {Yale},
  \citenamefont {Heremans}, \citenamefont {Zhou}, \citenamefont {Auer},
  \citenamefont {Burkard},\ and\ \citenamefont
  {Awschalom}}]{2016_Yale_NatPhoto}%
  \BibitemOpen
  \bibfield  {author} {\bibinfo {author} {\bibfnamefont {C.}~\bibnamefont
  {Yale}}, \bibinfo {author} {\bibfnamefont {F.}~\bibnamefont {Heremans}},
  \bibinfo {author} {\bibfnamefont {B.}~\bibnamefont {Zhou}}, \bibinfo {author}
  {\bibfnamefont {A.}~\bibnamefont {Auer}}, \bibinfo {author} {\bibfnamefont
  {G.}~\bibnamefont {Burkard}}, \ and\ \bibinfo {author} {\bibfnamefont
  {D.}~\bibnamefont {Awschalom}},\ }\href
  {https://doi.org/10.1038/nphoton.2015.278} {\bibfield  {journal} {\bibinfo
  {journal} {Nat. Photonics}\ }\textbf {\bibinfo {volume} {10}},\ \bibinfo
  {pages} {184} (\bibinfo {year} {2016})}\BibitemShut {NoStop}%
\bibitem [{\citenamefont {Sekiguchi}\ \emph {et~al.}(2017)\citenamefont
  {Sekiguchi}, \citenamefont {Niikura}, \citenamefont {Kuroiwa}, \citenamefont
  {Kano},\ and\ \citenamefont {Kosaka}}]{2017_Hideo_Nat_photo}%
  \BibitemOpen
  \bibfield  {author} {\bibinfo {author} {\bibfnamefont {Y.}~\bibnamefont
  {Sekiguchi}}, \bibinfo {author} {\bibfnamefont {N.}~\bibnamefont {Niikura}},
  \bibinfo {author} {\bibfnamefont {R.}~\bibnamefont {Kuroiwa}}, \bibinfo
  {author} {\bibfnamefont {H.}~\bibnamefont {Kano}}, \ and\ \bibinfo {author}
  {\bibfnamefont {H.}~\bibnamefont {Kosaka}},\ }\href {\doibase
  10.1038/nphoton.2017.40} {\bibfield  {journal} {\bibinfo  {journal} {Nat.
  Photonics}\ }\textbf {\bibinfo {volume} {11}},\ \bibinfo {pages} {309}
  (\bibinfo {year} {2017})}\BibitemShut {NoStop}%
\bibitem [{\citenamefont {Kleißler}\ \emph {et~al.}(2018)\citenamefont
  {Kleißler}, \citenamefont {Lazariev},\ and\ \citenamefont
  {Arroyo-Camejo}}]{npj_2018}%
  \BibitemOpen
  \bibfield  {author} {\bibinfo {author} {\bibfnamefont {F.}~\bibnamefont
  {Kleißler}}, \bibinfo {author} {\bibfnamefont {A.}~\bibnamefont {Lazariev}},
  \ and\ \bibinfo {author} {\bibfnamefont {S.}~\bibnamefont {Arroyo-Camejo}},\
  }\href {\doibase 10.1038/s41534-018-0098-7} {\bibfield  {journal} {\bibinfo
  {journal} {npj Quantum Inform.}\ }\textbf {\bibinfo {volume} {4}},\ \bibinfo
  {pages} {49} (\bibinfo {year} {2018})}\BibitemShut {NoStop}%
\bibitem [{\citenamefont {Huang}\ \emph {et~al.}(2019)\citenamefont {Huang},
  \citenamefont {Wu}, \citenamefont {Wang}, \citenamefont {Hou}, \citenamefont
  {Wang}, \citenamefont {Zhang}, \citenamefont {Lian}, \citenamefont {Liu},
  \citenamefont {Wang}, \citenamefont {Zhang}, \citenamefont {He},
  \citenamefont {Chang}, \citenamefont {Xu},\ and\ \citenamefont
  {Duan}}]{duan_prl_2019}%
  \BibitemOpen
  \bibfield  {author} {\bibinfo {author} {\bibfnamefont {Y.-Y.}\ \bibnamefont
  {Huang}}, \bibinfo {author} {\bibfnamefont {Y.-K.}\ \bibnamefont {Wu}},
  \bibinfo {author} {\bibfnamefont {F.}~\bibnamefont {Wang}}, \bibinfo {author}
  {\bibfnamefont {P.-Y.}\ \bibnamefont {Hou}}, \bibinfo {author} {\bibfnamefont
  {W.-B.}\ \bibnamefont {Wang}}, \bibinfo {author} {\bibfnamefont {W.-G.}\
  \bibnamefont {Zhang}}, \bibinfo {author} {\bibfnamefont {W.-Q.}\ \bibnamefont
  {Lian}}, \bibinfo {author} {\bibfnamefont {Y.-Q.}\ \bibnamefont {Liu}},
  \bibinfo {author} {\bibfnamefont {H.-Y.}\ \bibnamefont {Wang}}, \bibinfo
  {author} {\bibfnamefont {H.-Y.}\ \bibnamefont {Zhang}}, \bibinfo {author}
  {\bibfnamefont {L.}~\bibnamefont {He}}, \bibinfo {author} {\bibfnamefont
  {X.-Y.}\ \bibnamefont {Chang}}, \bibinfo {author} {\bibfnamefont
  {Y.}~\bibnamefont {Xu}}, \ and\ \bibinfo {author} {\bibfnamefont {L.-M.}\
  \bibnamefont {Duan}},\ }\href {\doibase 10.1103/PhysRevLett.122.010503}
  {\bibfield  {journal} {\bibinfo  {journal} {Phys. Rev. Lett.}\ }\textbf
  {\bibinfo {volume} {122}},\ \bibinfo {pages} {010503} (\bibinfo {year}
  {2019})}\BibitemShut {NoStop}%
\bibitem [{\citenamefont {Zhao}\ \emph {et~al.}(2017)\citenamefont {Zhao},
  \citenamefont {Cui}, \citenamefont {Xu}, \citenamefont {Sjöqvist},\ and\
  \citenamefont {Tong}}]{Zhao_PRA_atom_2017}%
  \BibitemOpen
  \bibfield  {author} {\bibinfo {author} {\bibfnamefont {P.}~\bibnamefont
  {Zhao}}, \bibinfo {author} {\bibfnamefont {X.-D.}\ \bibnamefont {Cui}},
  \bibinfo {author} {\bibfnamefont {G.}~\bibnamefont {Xu}}, \bibinfo {author}
  {\bibfnamefont {E.}~\bibnamefont {Sjöqvist}}, \ and\ \bibinfo {author}
  {\bibfnamefont {D.}~\bibnamefont {Tong}},\ }\href {\doibase
  10.1103/PhysRevA.96.052316} {\bibfield  {journal} {\bibinfo  {journal} {Phys.
  Rev. A}\ }\textbf {\bibinfo {volume} {96}},\ \bibinfo {pages} {052316}
  (\bibinfo {year} {2017})}\BibitemShut {NoStop}%
\bibitem [{\citenamefont {Liao}\ \emph {et~al.}(2019)\citenamefont {Liao},
  \citenamefont {Liu}, \citenamefont {Li},\ and\ \citenamefont
  {Du}}]{OL_atom_2019}%
  \BibitemOpen
  \bibfield  {author} {\bibinfo {author} {\bibfnamefont {K.-Y.}\ \bibnamefont
  {Liao}}, \bibinfo {author} {\bibfnamefont {X.-H.}\ \bibnamefont {Liu}},
  \bibinfo {author} {\bibfnamefont {Z.}~\bibnamefont {Li}}, \ and\ \bibinfo
  {author} {\bibfnamefont {Y.-X.}\ \bibnamefont {Du}},\ }\href {\doibase
  10.1364/OL.44.004801} {\bibfield  {journal} {\bibinfo  {journal} {Opt.
  Lett.}\ }\textbf {\bibinfo {volume} {44}},\ \bibinfo {pages} {4801} (\bibinfo
  {year} {2019})}\BibitemShut {NoStop}%
\bibitem [{\citenamefont {Liu}\ \emph {et~al.}(2020)\citenamefont {Liu},
  \citenamefont {Su},\ and\ \citenamefont {Yung}}]{Liu_PRR_atom_2020}%
  \BibitemOpen
  \bibfield  {author} {\bibinfo {author} {\bibfnamefont {B.-J.}\ \bibnamefont
  {Liu}}, \bibinfo {author} {\bibfnamefont {S.-L.}\ \bibnamefont {Su}}, \ and\
  \bibinfo {author} {\bibfnamefont {M.-H.}\ \bibnamefont {Yung}},\ }\href
  {\doibase 10.1103/PhysRevResearch.2.043130} {\bibfield  {journal} {\bibinfo
  {journal} {Phys. Rev. Research}\ }\textbf {\bibinfo {volume} {2}},\ \bibinfo
  {pages} {043130} (\bibinfo {year} {2020})}\BibitemShut {NoStop}%
\bibitem [{\citenamefont {Duan}\ \emph {et~al.}(2001)\citenamefont {Duan},
  \citenamefont {Cirac},\ and\ \citenamefont {Zoller}}]{Duan_science}%
  \BibitemOpen
  \bibfield  {author} {\bibinfo {author} {\bibfnamefont {L.-M.}\ \bibnamefont
  {Duan}}, \bibinfo {author} {\bibfnamefont {J.~I.}\ \bibnamefont {Cirac}}, \
  and\ \bibinfo {author} {\bibfnamefont {P.}~\bibnamefont {Zoller}},\ }\href
  {\doibase 10.1126/science.1058835} {\bibfield  {journal} {\bibinfo  {journal}
  {Science}\ }\textbf {\bibinfo {volume} {292}},\ \bibinfo {pages} {1695}
  (\bibinfo {year} {2001})}\BibitemShut {NoStop}%
\bibitem [{\citenamefont {Toyoda}\ \emph
  {et~al.}(2013{\natexlab{a}})\citenamefont {Toyoda}, \citenamefont {Uchida},
  \citenamefont {Noguchi}, \citenamefont {Haze},\ and\ \citenamefont
  {Urabe}}]{PhysRevA.87.052307}%
  \BibitemOpen
  \bibfield  {author} {\bibinfo {author} {\bibfnamefont {K.}~\bibnamefont
  {Toyoda}}, \bibinfo {author} {\bibfnamefont {K.}~\bibnamefont {Uchida}},
  \bibinfo {author} {\bibfnamefont {A.}~\bibnamefont {Noguchi}}, \bibinfo
  {author} {\bibfnamefont {S.}~\bibnamefont {Haze}}, \ and\ \bibinfo {author}
  {\bibfnamefont {S.}~\bibnamefont {Urabe}},\ }\href {\doibase
  10.1103/PhysRevA.87.052307} {\bibfield  {journal} {\bibinfo  {journal} {Phys.
  Rev. A}\ }\textbf {\bibinfo {volume} {87}},\ \bibinfo {pages} {052307}
  (\bibinfo {year} {2013}{\natexlab{a}})}\BibitemShut {NoStop}%
\bibitem [{\citenamefont {Leibfried}\ \emph {et~al.}(2003)\citenamefont
  {Leibfried}, \citenamefont {DeMarco}, \citenamefont {Meyer}, \citenamefont
  {Lucas}, \citenamefont {Barrett}, \citenamefont {Britton}, \citenamefont
  {Itano}, \citenamefont {Jelenković}, \citenamefont {Langer}, \citenamefont
  {Rosenband},\ and\ \citenamefont {Wineland}}]{leibfried_experimental_2003}%
  \BibitemOpen
  \bibfield  {author} {\bibinfo {author} {\bibfnamefont {D.}~\bibnamefont
  {Leibfried}}, \bibinfo {author} {\bibfnamefont {B.}~\bibnamefont {DeMarco}},
  \bibinfo {author} {\bibfnamefont {V.}~\bibnamefont {Meyer}}, \bibinfo
  {author} {\bibfnamefont {D.}~\bibnamefont {Lucas}}, \bibinfo {author}
  {\bibfnamefont {M.}~\bibnamefont {Barrett}}, \bibinfo {author} {\bibfnamefont
  {J.}~\bibnamefont {Britton}}, \bibinfo {author} {\bibfnamefont {W.~M.}\
  \bibnamefont {Itano}}, \bibinfo {author} {\bibfnamefont {B.}~\bibnamefont
  {Jelenković}}, \bibinfo {author} {\bibfnamefont {C.}~\bibnamefont {Langer}},
  \bibinfo {author} {\bibfnamefont {T.}~\bibnamefont {Rosenband}}, \ and\
  \bibinfo {author} {\bibfnamefont {D.~J.}\ \bibnamefont {Wineland}},\ }\href
  {\doibase 10.1038/nature01492} {\bibfield  {journal} {\bibinfo  {journal}
  {Nature (London)}\ }\textbf {\bibinfo {volume} {422}},\ \bibinfo {pages}
  {412} (\bibinfo {year} {2003})}\BibitemShut {NoStop}%
\bibitem [{\citenamefont {Ai}\ \emph {et~al.}(2020)\citenamefont {Ai},
  \citenamefont {Li}, \citenamefont {Hou}, \citenamefont {He}, \citenamefont
  {Qian}, \citenamefont {Xue}, \citenamefont {Cui}, \citenamefont {Huang},
  \citenamefont {Li},\ and\ \citenamefont {Guo}}]{PhysRevApplied.14.054062}%
  \BibitemOpen
  \bibfield  {author} {\bibinfo {author} {\bibfnamefont {M.-Z.}\ \bibnamefont
  {Ai}}, \bibinfo {author} {\bibfnamefont {S.}~\bibnamefont {Li}}, \bibinfo
  {author} {\bibfnamefont {Z.}~\bibnamefont {Hou}}, \bibinfo {author}
  {\bibfnamefont {R.}~\bibnamefont {He}}, \bibinfo {author} {\bibfnamefont
  {Z.-H.}\ \bibnamefont {Qian}}, \bibinfo {author} {\bibfnamefont {Z.-Y.}\
  \bibnamefont {Xue}}, \bibinfo {author} {\bibfnamefont {J.-M.}\ \bibnamefont
  {Cui}}, \bibinfo {author} {\bibfnamefont {Y.-F.}\ \bibnamefont {Huang}},
  \bibinfo {author} {\bibfnamefont {C.-F.}\ \bibnamefont {Li}}, \ and\ \bibinfo
  {author} {\bibfnamefont {G.-C.}\ \bibnamefont {Guo}},\ }\href {\doibase
  10.1103/PhysRevApplied.14.054062} {\bibfield  {journal} {\bibinfo  {journal}
  {Phys. Rev. Applied}\ }\textbf {\bibinfo {volume} {14}},\ \bibinfo {pages}
  {054062} (\bibinfo {year} {2020})}\BibitemShut {NoStop}%
\bibitem [{\citenamefont {Zhang}\ \emph {et~al.}(2021)\citenamefont {Zhang},
  \citenamefont {Yan}, \citenamefont {Li}, \citenamefont {Ding}, \citenamefont
  {Bu}, \citenamefont {Chen}, \citenamefont {Su}, \citenamefont {Zhou},\ and\
  \citenamefont {Feng}}]{PhysRevLett.127.030502}%
  \BibitemOpen
  \bibfield  {author} {\bibinfo {author} {\bibfnamefont {J.}~\bibnamefont
  {Zhang}}, \bibinfo {author} {\bibfnamefont {L.}~\bibnamefont {Yan}}, \bibinfo
  {author} {\bibfnamefont {J.}~\bibnamefont {Li}}, \bibinfo {author}
  {\bibfnamefont {G.}~\bibnamefont {Ding}}, \bibinfo {author} {\bibfnamefont
  {J.}~\bibnamefont {Bu}}, \bibinfo {author} {\bibfnamefont {L.}~\bibnamefont
  {Chen}}, \bibinfo {author} {\bibfnamefont {S.}~\bibnamefont {Su}}, \bibinfo
  {author} {\bibfnamefont {F.}~\bibnamefont {Zhou}}, \ and\ \bibinfo {author}
  {\bibfnamefont {M.}~\bibnamefont {Feng}},\ }\href {\doibase
  10.1103/PhysRevLett.127.030502} {\bibfield  {journal} {\bibinfo  {journal}
  {Phys. Rev. Lett.}\ }\textbf {\bibinfo {volume} {127}},\ \bibinfo {pages}
  {030502} (\bibinfo {year} {2021})}\BibitemShut {NoStop}%
\bibitem [{\citenamefont {Abdumalikov~Jr}\ \emph {et~al.}(2013)\citenamefont
  {Abdumalikov~Jr}, \citenamefont {Fink}, \citenamefont {Juliusson},
  \citenamefont {Pechal}, \citenamefont {Berger}, \citenamefont {Wallraff},\
  and\ \citenamefont {Filipp}}]{2013_Filipp_Nature}%
  \BibitemOpen
  \bibfield  {author} {\bibinfo {author} {\bibfnamefont {A.~A.}\ \bibnamefont
  {Abdumalikov~Jr}}, \bibinfo {author} {\bibfnamefont {J.~M.}\ \bibnamefont
  {Fink}}, \bibinfo {author} {\bibfnamefont {K.}~\bibnamefont {Juliusson}},
  \bibinfo {author} {\bibfnamefont {M.}~\bibnamefont {Pechal}}, \bibinfo
  {author} {\bibfnamefont {S.}~\bibnamefont {Berger}}, \bibinfo {author}
  {\bibfnamefont {A.}~\bibnamefont {Wallraff}}, \ and\ \bibinfo {author}
  {\bibfnamefont {S.}~\bibnamefont {Filipp}},\ }\href {\doibase
  10.1038/nature12010} {\bibfield  {journal} {\bibinfo  {journal} {Nature
  (London)}\ }\textbf {\bibinfo {volume} {496}},\ \bibinfo {pages} {482}
  (\bibinfo {year} {2013})}\BibitemShut {NoStop}%
\bibitem [{\citenamefont {Yan}\ \emph {et~al.}(2019)\citenamefont {Yan},
  \citenamefont {Liu}, \citenamefont {Xu}, \citenamefont {Song}, \citenamefont
  {Liu}, \citenamefont {Zhang}, \citenamefont {Deng}, \citenamefont {Yan},
  \citenamefont {Rong}, \citenamefont {Huang}, \citenamefont {Yung},
  \citenamefont {Chen},\ and\ \citenamefont {Yu}}]{2019_Yu_PRL}%
  \BibitemOpen
  \bibfield  {author} {\bibinfo {author} {\bibfnamefont {T.}~\bibnamefont
  {Yan}}, \bibinfo {author} {\bibfnamefont {B.-J.}\ \bibnamefont {Liu}},
  \bibinfo {author} {\bibfnamefont {K.}~\bibnamefont {Xu}}, \bibinfo {author}
  {\bibfnamefont {C.}~\bibnamefont {Song}}, \bibinfo {author} {\bibfnamefont
  {S.}~\bibnamefont {Liu}}, \bibinfo {author} {\bibfnamefont {Z.}~\bibnamefont
  {Zhang}}, \bibinfo {author} {\bibfnamefont {H.}~\bibnamefont {Deng}},
  \bibinfo {author} {\bibfnamefont {Z.}~\bibnamefont {Yan}}, \bibinfo {author}
  {\bibfnamefont {H.}~\bibnamefont {Rong}}, \bibinfo {author} {\bibfnamefont
  {K.}~\bibnamefont {Huang}}, \bibinfo {author} {\bibfnamefont {M.-H.}\
  \bibnamefont {Yung}}, \bibinfo {author} {\bibfnamefont {Y.}~\bibnamefont
  {Chen}}, \ and\ \bibinfo {author} {\bibfnamefont {D.}~\bibnamefont {Yu}},\
  }\href {\doibase 10.1103/PhysRevLett.122.080501} {\bibfield  {journal}
  {\bibinfo  {journal} {Phys. Rev. Lett.}\ }\textbf {\bibinfo {volume} {122}},\
  \bibinfo {pages} {080501} (\bibinfo {year} {2019})}\BibitemShut {NoStop}%
\bibitem [{\citenamefont {Xu}\ \emph {et~al.}(2018)\citenamefont {Xu},
  \citenamefont {Cai}, \citenamefont {Ma}, \citenamefont {Mu}, \citenamefont
  {Hu}, \citenamefont {Chen}, \citenamefont {Wang}, \citenamefont {Song},
  \citenamefont {Xue}, \citenamefont {Yin},\ and\ \citenamefont
  {Sun}}]{PhysRevLett.121.110501}%
  \BibitemOpen
  \bibfield  {author} {\bibinfo {author} {\bibfnamefont {Y.}~\bibnamefont
  {Xu}}, \bibinfo {author} {\bibfnamefont {W.}~\bibnamefont {Cai}}, \bibinfo
  {author} {\bibfnamefont {Y.}~\bibnamefont {Ma}}, \bibinfo {author}
  {\bibfnamefont {X.}~\bibnamefont {Mu}}, \bibinfo {author} {\bibfnamefont
  {L.}~\bibnamefont {Hu}}, \bibinfo {author} {\bibfnamefont {T.}~\bibnamefont
  {Chen}}, \bibinfo {author} {\bibfnamefont {H.}~\bibnamefont {Wang}}, \bibinfo
  {author} {\bibfnamefont {Y.}~\bibnamefont {Song}}, \bibinfo {author}
  {\bibfnamefont {Z.-Y.}\ \bibnamefont {Xue}}, \bibinfo {author} {\bibfnamefont
  {Z.-q.}\ \bibnamefont {Yin}}, \ and\ \bibinfo {author} {\bibfnamefont
  {L.}~\bibnamefont {Sun}},\ }\href {\doibase 10.1103/PhysRevLett.121.110501}
  {\bibfield  {journal} {\bibinfo  {journal} {Phys. Rev. Lett.}\ }\textbf
  {\bibinfo {volume} {121}},\ \bibinfo {pages} {110501} (\bibinfo {year}
  {2018})}\BibitemShut {NoStop}%
\bibitem [{\citenamefont {Xu}\ \emph {et~al.}(2020)\citenamefont {Xu},
  \citenamefont {Hua}, \citenamefont {Chen}, \citenamefont {Pan}, \citenamefont
  {Li}, \citenamefont {Han}, \citenamefont {Cai}, \citenamefont {Ma},
  \citenamefont {Wang}, \citenamefont {Song}, \citenamefont {Xue},\ and\
  \citenamefont {Sun}}]{PhysRevLett.124.230503}%
  \BibitemOpen
  \bibfield  {author} {\bibinfo {author} {\bibfnamefont {Y.}~\bibnamefont
  {Xu}}, \bibinfo {author} {\bibfnamefont {Z.}~\bibnamefont {Hua}}, \bibinfo
  {author} {\bibfnamefont {T.}~\bibnamefont {Chen}}, \bibinfo {author}
  {\bibfnamefont {X.}~\bibnamefont {Pan}}, \bibinfo {author} {\bibfnamefont
  {X.}~\bibnamefont {Li}}, \bibinfo {author} {\bibfnamefont {J.}~\bibnamefont
  {Han}}, \bibinfo {author} {\bibfnamefont {W.}~\bibnamefont {Cai}}, \bibinfo
  {author} {\bibfnamefont {Y.}~\bibnamefont {Ma}}, \bibinfo {author}
  {\bibfnamefont {H.}~\bibnamefont {Wang}}, \bibinfo {author} {\bibfnamefont
  {Y.}~\bibnamefont {Song}}, \bibinfo {author} {\bibfnamefont {Z.-Y.}\
  \bibnamefont {Xue}}, \ and\ \bibinfo {author} {\bibfnamefont
  {L.}~\bibnamefont {Sun}},\ }\href {\doibase 10.1103/PhysRevLett.124.230503}
  {\bibfield  {journal} {\bibinfo  {journal} {Phys. Rev. Lett.}\ }\textbf
  {\bibinfo {volume} {124}},\ \bibinfo {pages} {230503} (\bibinfo {year}
  {2020})}\BibitemShut {NoStop}%
\bibitem [{\citenamefont {Zhang}\ \emph {et~al.}(2019)\citenamefont {Zhang},
  \citenamefont {Zhao}, \citenamefont {Wang}, \citenamefont {Xiang},
  \citenamefont {Jia}, \citenamefont {Duan}, \citenamefont {Tong},
  \citenamefont {Yin},\ and\ \citenamefont {Guo}}]{Zhang_2019}%
  \BibitemOpen
  \bibfield  {author} {\bibinfo {author} {\bibfnamefont {Z.}~\bibnamefont
  {Zhang}}, \bibinfo {author} {\bibfnamefont {P.}~\bibnamefont {Zhao}},
  \bibinfo {author} {\bibfnamefont {T.}~\bibnamefont {Wang}}, \bibinfo {author}
  {\bibfnamefont {L.}~\bibnamefont {Xiang}}, \bibinfo {author} {\bibfnamefont
  {Z.}~\bibnamefont {Jia}}, \bibinfo {author} {\bibfnamefont {P.}~\bibnamefont
  {Duan}}, \bibinfo {author} {\bibfnamefont {D.}~\bibnamefont {Tong}}, \bibinfo
  {author} {\bibfnamefont {Y.}~\bibnamefont {Yin}}, \ and\ \bibinfo {author}
  {\bibfnamefont {G.}~\bibnamefont {Guo}},\ }\href {\doibase
  10.1088/1367-2630/ab2e26} {\bibfield  {journal} {\bibinfo  {journal} {New J.
  Phys.}\ }\textbf {\bibinfo {volume} {21}},\ \bibinfo {pages} {073024}
  (\bibinfo {year} {2019})}\BibitemShut {NoStop}%
\bibitem [{\citenamefont {Zhao}\ \emph {et~al.}(2021)\citenamefont {Zhao},
  \citenamefont {Dong}, \citenamefont {Zhang}, \citenamefont {Guo},
  \citenamefont {Tong},\ and\ \citenamefont {Yin}}]{Zhao_2021}%
  \BibitemOpen
  \bibfield  {author} {\bibinfo {author} {\bibfnamefont {P.}~\bibnamefont
  {Zhao}}, \bibinfo {author} {\bibfnamefont {Z.}~\bibnamefont {Dong}}, \bibinfo
  {author} {\bibfnamefont {Z.}~\bibnamefont {Zhang}}, \bibinfo {author}
  {\bibfnamefont {G.}~\bibnamefont {Guo}}, \bibinfo {author} {\bibfnamefont
  {D.}~\bibnamefont {Tong}}, \ and\ \bibinfo {author} {\bibfnamefont
  {Y.}~\bibnamefont {Yin}},\ }\href {\doibase 10.1007/s11433-020-1641-1}
  {\bibfield  {journal} {\bibinfo  {journal} {Sci. China-Phys. Mech. Astron.}\
  }\textbf {\bibinfo {volume} {64}},\ \bibinfo {pages} {250362} (\bibinfo
  {year} {2021})}\BibitemShut {NoStop}%
\bibitem [{\citenamefont {Boykin}\ \emph {et~al.}(1999)\citenamefont {Boykin},
  \citenamefont {Mor}, \citenamefont {Pulver}, \citenamefont {Roychowdhury},\
  and\ \citenamefont {Vatan}}]{1999_Boykin_universal}%
  \BibitemOpen
  \bibfield  {author} {\bibinfo {author} {\bibfnamefont {P.}~\bibnamefont
  {Boykin}}, \bibinfo {author} {\bibfnamefont {T.}~\bibnamefont {Mor}},
  \bibinfo {author} {\bibfnamefont {M.}~\bibnamefont {Pulver}}, \bibinfo
  {author} {\bibfnamefont {V.}~\bibnamefont {Roychowdhury}}, \ and\ \bibinfo
  {author} {\bibfnamefont {F.}~\bibnamefont {Vatan}},\ }\href
  {https://arxiv.org/abs/quant-ph/9906054} {\bibfield  {journal} {\bibinfo
  {journal} {arXiv preprint quant-ph/9906054}\ } (\bibinfo {year}
  {1999})}\BibitemShut {NoStop}%
\bibitem [{\citenamefont {Nielsen}\ and\ \citenamefont
  {Chuang}(2010)}]{nielsen_chuang_2010}%
  \BibitemOpen
  \bibfield  {author} {\bibinfo {author} {\bibfnamefont {M.}~\bibnamefont
  {Nielsen}}\ and\ \bibinfo {author} {\bibfnamefont {I.}~\bibnamefont
  {Chuang}},\ }\href {\doibase 10.1017/CBO9780511976667} {\emph {\bibinfo
  {title} {Quantum Computation and Quantum Information: 10th Anniversary
  Edition}}}\ (\bibinfo  {publisher} {Cambridge University Press},\ \bibinfo
  {year} {2010})\BibitemShut {NoStop}%
\bibitem [{\citenamefont {Howard}\ \emph {et~al.}(2006)\citenamefont {Howard},
  \citenamefont {Twamley}, \citenamefont {Wittmann}, \citenamefont {Gaebel},
  \citenamefont {Jelezko},\ and\ \citenamefont {Wrachtrup}}]{Howard_2006}%
  \BibitemOpen
  \bibfield  {author} {\bibinfo {author} {\bibfnamefont {M.}~\bibnamefont
  {Howard}}, \bibinfo {author} {\bibfnamefont {J.}~\bibnamefont {Twamley}},
  \bibinfo {author} {\bibfnamefont {C.}~\bibnamefont {Wittmann}}, \bibinfo
  {author} {\bibfnamefont {T.}~\bibnamefont {Gaebel}}, \bibinfo {author}
  {\bibfnamefont {F.}~\bibnamefont {Jelezko}}, \ and\ \bibinfo {author}
  {\bibfnamefont {J.}~\bibnamefont {Wrachtrup}},\ }\href {\doibase
  10.1088/1367-2630/8/3/033} {\bibfield  {journal} {\bibinfo  {journal} {New J.
  Phys.}\ }\textbf {\bibinfo {volume} {8}},\ \bibinfo {pages} {33} (\bibinfo
  {year} {2006})}\BibitemShut {NoStop}%
\bibitem [{\citenamefont {Toyoda}\ \emph
  {et~al.}(2013{\natexlab{b}})\citenamefont {Toyoda}, \citenamefont {Uchida},
  \citenamefont {Noguchi}, \citenamefont {Haze},\ and\ \citenamefont
  {Urabe}}]{2013_Urabe_Holonomic}%
  \BibitemOpen
  \bibfield  {author} {\bibinfo {author} {\bibfnamefont {K.}~\bibnamefont
  {Toyoda}}, \bibinfo {author} {\bibfnamefont {K.}~\bibnamefont {Uchida}},
  \bibinfo {author} {\bibfnamefont {A.}~\bibnamefont {Noguchi}}, \bibinfo
  {author} {\bibfnamefont {S.}~\bibnamefont {Haze}}, \ and\ \bibinfo {author}
  {\bibfnamefont {S.}~\bibnamefont {Urabe}},\ }\href {\doibase
  10.1103/PhysRevA.87.052307} {\bibfield  {journal} {\bibinfo  {journal} {Phys.
  Rev. A}\ }\textbf {\bibinfo {volume} {87}},\ \bibinfo {pages} {052307}
  (\bibinfo {year} {2013}{\natexlab{b}})}\BibitemShut {NoStop}%
\bibitem [{\citenamefont {Liu}\ \emph {et~al.}(2019)\citenamefont {Liu},
  \citenamefont {Song}, \citenamefont {Xue}, \citenamefont {Wang},\ and\
  \citenamefont {Yung}}]{Xuezhengyuan_2019}%
  \BibitemOpen
  \bibfield  {author} {\bibinfo {author} {\bibfnamefont {B.-J.}\ \bibnamefont
  {Liu}}, \bibinfo {author} {\bibfnamefont {X.-K.}\ \bibnamefont {Song}},
  \bibinfo {author} {\bibfnamefont {Z.-Y.}\ \bibnamefont {Xue}}, \bibinfo
  {author} {\bibfnamefont {X.}~\bibnamefont {Wang}}, \ and\ \bibinfo {author}
  {\bibfnamefont {M.-H.}\ \bibnamefont {Yung}},\ }\href {\doibase
  10.1103/PhysRevLett.123.100501} {\bibfield  {journal} {\bibinfo  {journal}
  {Phys. Rev. Lett.}\ }\textbf {\bibinfo {volume} {123}},\ \bibinfo {pages}
  {100501} (\bibinfo {year} {2019})}\BibitemShut {NoStop}%
\bibitem [{\citenamefont {Song}\ \emph {et~al.}(2020)\citenamefont {Song},
  \citenamefont {Lim},\ and\ \citenamefont {Ahn}}]{2020_Ahn_PRR}%
  \BibitemOpen
  \bibfield  {author} {\bibinfo {author} {\bibfnamefont {Y.}~\bibnamefont
  {Song}}, \bibinfo {author} {\bibfnamefont {J.}~\bibnamefont {Lim}}, \ and\
  \bibinfo {author} {\bibfnamefont {J.}~\bibnamefont {Ahn}},\ }\href {\doibase
  10.1103/PhysRevResearch.2.023045} {\bibfield  {journal} {\bibinfo  {journal}
  {Phys. Rev. Research}\ }\textbf {\bibinfo {volume} {2}},\ \bibinfo {pages}
  {023045} (\bibinfo {year} {2020})}\BibitemShut {NoStop}%
\bibitem [{\citenamefont {Parker}\ \emph {et~al.}(2013)\citenamefont {Parker},
  \citenamefont {Ha},\ and\ \citenamefont {Chin}}]{2013_Chin_Shaking}%
  \BibitemOpen
  \bibfield  {author} {\bibinfo {author} {\bibfnamefont {C.}~\bibnamefont
  {Parker}}, \bibinfo {author} {\bibfnamefont {L.-C.}\ \bibnamefont {Ha}}, \
  and\ \bibinfo {author} {\bibfnamefont {C.}~\bibnamefont {Chin}},\ }\href
  {\doibase 10.1038/nphys2789} {\bibfield  {journal} {\bibinfo  {journal} {Nat.
  Phys.}\ }\textbf {\bibinfo {volume} {9}},\ \bibinfo {pages} {769} (\bibinfo
  {year} {2013})}\BibitemShut {NoStop}%
\bibitem [{\citenamefont {Hauke}\ \emph {et~al.}(2014)\citenamefont {Hauke},
  \citenamefont {Lewenstein},\ and\ \citenamefont {Eckardt}}]{2014_Hauke_PRL}%
  \BibitemOpen
  \bibfield  {author} {\bibinfo {author} {\bibfnamefont {P.}~\bibnamefont
  {Hauke}}, \bibinfo {author} {\bibfnamefont {M.}~\bibnamefont {Lewenstein}}, \
  and\ \bibinfo {author} {\bibfnamefont {A.}~\bibnamefont {Eckardt}},\ }\href
  {\doibase 10.1103/PhysRevLett.113.045303} {\bibfield  {journal} {\bibinfo
  {journal} {Phys. Rev. Lett.}\ }\textbf {\bibinfo {volume} {113}},\ \bibinfo
  {pages} {045303} (\bibinfo {year} {2014})}\BibitemShut {NoStop}%
\bibitem [{\citenamefont {Wang}\ \emph {et~al.}(2016)\citenamefont {Wang},
  \citenamefont {Kumar}, \citenamefont {Wu},\ and\ \citenamefont
  {Weiss}}]{wang2016single}%
  \BibitemOpen
  \bibfield  {author} {\bibinfo {author} {\bibfnamefont {Y.}~\bibnamefont
  {Wang}}, \bibinfo {author} {\bibfnamefont {A.}~\bibnamefont {Kumar}},
  \bibinfo {author} {\bibfnamefont {T.-Y.}\ \bibnamefont {Wu}}, \ and\ \bibinfo
  {author} {\bibfnamefont {D.}~\bibnamefont {Weiss}},\ }\href {\doibase
  10.1126/science.aaf2581} {\bibfield  {journal} {\bibinfo  {journal}
  {Science}\ }\textbf {\bibinfo {volume} {352}},\ \bibinfo {pages} {1562}
  (\bibinfo {year} {2016})}\BibitemShut {NoStop}%
\bibitem [{\citenamefont {Qiu}\ \emph {et~al.}(2020)\citenamefont {Qiu},
  \citenamefont {Zou}, \citenamefont {Qi},\ and\ \citenamefont
  {Li}}]{2020_Li_NPJ}%
  \BibitemOpen
  \bibfield  {author} {\bibinfo {author} {\bibfnamefont {X.}~\bibnamefont
  {Qiu}}, \bibinfo {author} {\bibfnamefont {J.}~\bibnamefont {Zou}}, \bibinfo
  {author} {\bibfnamefont {X.}~\bibnamefont {Qi}}, \ and\ \bibinfo {author}
  {\bibfnamefont {X.}~\bibnamefont {Li}},\ }\href {\doibase
  10.1038/s41534-020-00315-9} {\bibfield  {journal} {\bibinfo  {journal} {npj
  Quantum Inform.}\ }\textbf {\bibinfo {volume} {6}},\ \bibinfo {pages} {87}
  (\bibinfo {year} {2020})}\BibitemShut {NoStop}%
\bibitem [{\citenamefont {De~Chiara}\ \emph {et~al.}(2008)\citenamefont
  {De~Chiara}, \citenamefont {Calarco}, \citenamefont {Anderlini},
  \citenamefont {Montangero}, \citenamefont {Lee}, \citenamefont {Brown},
  \citenamefont {Phillips},\ and\ \citenamefont {Porto}}]{2008_Porto_PRA}%
  \BibitemOpen
  \bibfield  {author} {\bibinfo {author} {\bibfnamefont {G.}~\bibnamefont
  {De~Chiara}}, \bibinfo {author} {\bibfnamefont {T.}~\bibnamefont {Calarco}},
  \bibinfo {author} {\bibfnamefont {M.}~\bibnamefont {Anderlini}}, \bibinfo
  {author} {\bibfnamefont {S.}~\bibnamefont {Montangero}}, \bibinfo {author}
  {\bibfnamefont {P.}~\bibnamefont {Lee}}, \bibinfo {author} {\bibfnamefont
  {B.}~\bibnamefont {Brown}}, \bibinfo {author} {\bibfnamefont
  {W.}~\bibnamefont {Phillips}}, \ and\ \bibinfo {author} {\bibfnamefont
  {J.}~\bibnamefont {Porto}},\ }\href {\doibase 10.1103/PhysRevA.77.052333}
  {\bibfield  {journal} {\bibinfo  {journal} {Phys. Rev. A}\ }\textbf {\bibinfo
  {volume} {77}},\ \bibinfo {pages} {052333} (\bibinfo {year}
  {2008})}\BibitemShut {NoStop}%
\bibitem [{\citenamefont {Jensen}\ \emph {et~al.}(2019)\citenamefont {Jensen},
  \citenamefont {Sørensen}, \citenamefont {Mølmer},\ and\ \citenamefont
  {Sherson}}]{2019_Sherson_PRA}%
  \BibitemOpen
  \bibfield  {author} {\bibinfo {author} {\bibfnamefont {J.}~\bibnamefont
  {Jensen}}, \bibinfo {author} {\bibfnamefont {J.}~\bibnamefont {Sørensen}},
  \bibinfo {author} {\bibfnamefont {K.}~\bibnamefont {Mølmer}}, \ and\
  \bibinfo {author} {\bibfnamefont {J.}~\bibnamefont {Sherson}},\ }\href
  {\doibase 10.1103/PhysRevA.100.052314} {\bibfield  {journal} {\bibinfo
  {journal} {Phys. Rev. A}\ }\textbf {\bibinfo {volume} {100}},\ \bibinfo
  {pages} {052314} (\bibinfo {year} {2019})}\BibitemShut {NoStop}%
\bibitem [{\citenamefont {Mamaev}\ \emph {et~al.}(2020)\citenamefont {Mamaev},
  \citenamefont {Thywissen},\ and\ \citenamefont {Rey}}]{2020_Rey_AQT}%
  \BibitemOpen
  \bibfield  {author} {\bibinfo {author} {\bibfnamefont {M.}~\bibnamefont
  {Mamaev}}, \bibinfo {author} {\bibfnamefont {J.~H.}\ \bibnamefont
  {Thywissen}}, \ and\ \bibinfo {author} {\bibfnamefont {A.~M.}\ \bibnamefont
  {Rey}},\ }\href {\doibase https://doi.org/10.1002/qute.201900132} {\bibfield
  {journal} {\bibinfo  {journal} {Advanced Quantum Technologies}\ }\textbf
  {\bibinfo {volume} {3}},\ \bibinfo {pages} {1900132} (\bibinfo {year}
  {2020})}\BibitemShut {NoStop}%
\bibitem [{\citenamefont {{Hartke}}\ \emph {et~al.}(2021)\citenamefont
  {{Hartke}}, \citenamefont {{Oreg}}, \citenamefont {{Jia}},\ and\
  \citenamefont {{Zwierlein}}}]{2021_Zwierlein_arXiv}%
  \BibitemOpen
  \bibfield  {author} {\bibinfo {author} {\bibfnamefont {T.}~\bibnamefont
  {{Hartke}}}, \bibinfo {author} {\bibfnamefont {B.}~\bibnamefont {{Oreg}}},
  \bibinfo {author} {\bibfnamefont {N.}~\bibnamefont {{Jia}}}, \ and\ \bibinfo
  {author} {\bibfnamefont {M.}~\bibnamefont {{Zwierlein}}},\ }\href@noop {}
  {\bibfield  {journal} {\bibinfo  {journal} {arXiv e-prints}\ ,\ \bibinfo
  {eid} {arXiv:2103.13992}} (\bibinfo {year} {2021})},\ \Eprint
  {http://arxiv.org/abs/2103.13992} {arXiv:2103.13992 [cond-mat.quant-gas]}
  \BibitemShut {NoStop}%
\bibitem [{\citenamefont {Nielsen}\ and\ \citenamefont
  {Chuang}(2002)}]{nielsen2002quantum}%
  \BibitemOpen
  \bibfield  {author} {\bibinfo {author} {\bibfnamefont {M.~A.}\ \bibnamefont
  {Nielsen}}\ and\ \bibinfo {author} {\bibfnamefont {I.}~\bibnamefont
  {Chuang}},\ }\href@noop {} {\enquote {\bibinfo {title} {Quantum computation
  and quantum information},}\ } (\bibinfo {year} {2002})\BibitemShut {NoStop}%
\end{thebibliography}%

\bibliographystyle{apsrev4-1}

\begin{widetext} 
\newpage

\renewcommand{\theequation}{S\arabic{equation}}
\renewcommand{\thesection}{S-\arabic{section}}
\renewcommand{\thefigure}{S\arabic{figure}}
\renewcommand{\thetable}{S\arabic{table}}
\setcounter{equation}{0}
\setcounter{figure}{0}
\setcounter{table}{0}

\begin{center}
{\Huge \bf Supplementary Material} \\
\end{center}

\section{Experimental lattice setup}

{
Our experiment is based on a Bose-Einstein condensate (BEC) of  $^{87}$Rb atoms confined in a harmonic trap, with trapping frequencies 
[$(\omega_ x, \omega_y, \omega_z) = 2\pi \times (24 {\rm Hz}, 48 {\rm Hz}, 55 {\rm Hz})$] 
along the three spatial directions. The BEC is loaded into a one-dimensional lattice formed by counter-propagating laser beams along the $x$ direction.} 
To achieve the holonomic quantum gate manipulation on the atom-orbital qubit, on the one hand, the lattice laser is required to be rapidly turned on and off in order to form square wave pulses in the state preparation, on the other hand, the variation of lattice depth must accurately follow the sinusoidal form in the gate operation. Our experimental setup is carefully designed with two AOMs cooperating (Fig.~\ref{fig:s1}) to improve control precision.
The AOM1 works with the laser power feedback control loop, consisting of PD (Photodetector), AWG1 (Arbitrary wave generator),
PI (Proportional-Integral) feedback circuit, and variable power RF (Radio frequency) source1. A voltage signal in PD for programming the lattice depth $V(t)=V_0+\Delta V(t)$ is achieved by the diffracted light of AOM1, where the time-dependent part $\Delta V(t)$ corresponds to a sinusoidal wave pulse.
AWG1 generates the control signal $V_{AWG1}\propto V(t)$. Both signals are connected into
PI circuit, and its output controls the power of RF source1 so that the diffraction efficiency of AOM1 is delicately adjusted in order to keep the value of $V_{AWG1}$ stable.  The AOM2
works with the fast control module, consisting of RF switch, RF source2 and AWG2 that generates square wave pulses.
According to RF source2, the RF switch can control AOM2 to realize the sharp variation of lattice depth corresponding to square wave pulses.
With this feedback control design using two AOMs, the rise and fall time is maintained below 100ns, the modulation amplitude uncertainty within 2$\%$ and the laser power fluctuation below $0.2\%@1$s.


\begin{figure}[htp]
\centering
\includegraphics[width=1\textwidth]{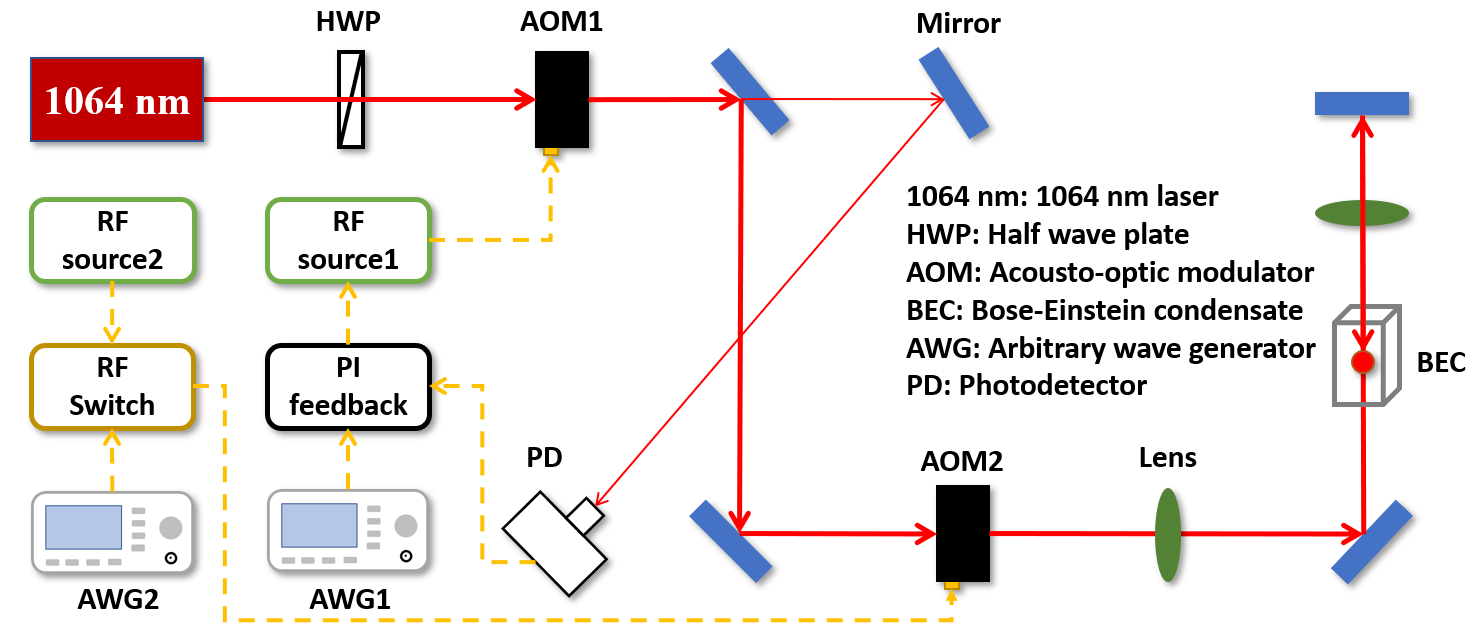}
\caption{Experimental set-up. The optical lattice is formed by 1064 nm laser, and its optical power (lattice depth) is controlled by two AOMs. The red solid line with arrow indicates the optical path and the yellow dashed line shows the direction of electrical signal transmission.
}
\label{fig:s1}
\end{figure}

\section{Qubit initialization}
We use a shortcut method~\cite{NJP_zhou} 
for atom-orbital qubit initialization. 
After preparation of an atomic Bose-Einstein condensate in a harmonic trap, we load the atoms directly  into the desired orbital state of the optical lattice.
For a given targeting quantum superposition state of 
$|s\rangle$ and $|d\rangle$ orbitals, a pulse sequence 
of the lattice-forming laser is designed to maximize the 
state preparation fidelity (see an example in Fig. 3(a)).
The shortcut method typically involves a few cycles of turn-on and off of the laser intensity, with a time duration $\tau_{{\rm on}, i}$, and $\tau_{{\rm off}, i}$ ($i$ labeling different cycles), respectively.
The time duration $\tau_{{\rm on}, i}$ and $\tau_{{\rm off}_i}$ defining the pulse sequence are variationally determined.
The pulse sequences for preparation in the initial states ($\mathbb{S} \equiv \{
\left\lvert 0\right\rangle ,
\left\lvert 1\right\rangle,
\left\lvert +\right\rangle,
\left\lvert -\right\rangle,
\left\lvert +i\right\rangle,
\left\lvert -i\right\rangle \}$) are provided in Table~\ref{tab:initial_seq}.
This shortcut method is demonstrated an efficient protocol for atom-orbital qubit initialization.
\begin{table}[htp]
\begin{tabular}{cccccccc}
\hline
{Initial state}          & {$\tau_{{\rm on},1}$}    & {$\tau_{{\rm off},1}$}    & {$\tau_{{\rm on}, 2}$}    & {$\tau_{{\rm off}, 2}$}    & {$\tau_{{\rm on},3}$}    & {$\tau_{{\rm off}, 3}$} & $(\mu s)$   \\\hline
{$|0\rangle$}             & {64.5}      & {40.7}       & {62.1}                & {41.4}                & {}                 & {}        & {}         \\
{$|1\rangle$}             & {28.1}      & {35.1}       & {40.6}                & {27.0}                & {}                 & {}        & {}        \\
{$|+\rangle$}             & {31.0}      & {7.7}       & {1.5}                & {41.1}                & {12.1}                 & {8.7}   & {}        \\
{$|-\rangle$}            & {38.4}       & {61.7}       & {38.6}                & {30.7}                & {}                  & {}       & {}           \\
{$|+i\rangle$}            & {52.0}       & {80.0}       & {3.0}                & {4.0}                & {}                  & {}        & {}         \\
{$|-i\rangle$}             & {23.3}      & {43.7}      & {56.5}                 & {41.8}                & {}                  & {}      & {}           \\
\hline
\end{tabular}
\caption{The laser pulse sequences for atom-orbital qubit initialization.}
\label{tab:initial_seq}
\end{table}

\section{Pulse sequence design}
In constructing the atom-orbital qubit and the holonomic quantum control, we have neglected atomic interactions, which is valid in a relatively shallower lattice. The requirement of having unharmonic band structures to avoid unwanted inter-band transitions also favors shallow lattice. At the same time, the lattice cannot be too shallow  for the corresponding coupling $\lambda$ would be too small to maintain the validity of  the rotating wave approximation as used in constructing the holonomic dynamical invariant (Eq.~(4)). Taking these constraints into account, we choose a moderate lattice depth of $V_0 = 5E_r$. We have also tried  other lattice depths in the experiment, and the choice of $5E_r$ is found to perform the best.

Through numerical tests, we observe that a single holonomic unitary operation given Eq.~(6) is not adequate to produce universal SU(2) quantum gates. To resolve this problem, we consider a concatenated unitary composed of a sequence of holonomic unitary operations,
\begin{equation}
U=U_{\beta_M \varphi_{M}}U_{\beta_{M-1} \varphi_{M-1}}...U_{\beta_2\varphi_{2} }U_{\beta_1\varphi_{1}}
\end{equation}
in order to construct generic quantum gates.
The control sequences ($\beta_j, \varphi_j$) are obtained by optimizing the gate fidelity,
\begin{equation}
F=| \operatorname{tr}\left(U^{\dagger} U_{\text { target }}\right) | /2,
\label{eq:suppfidelity}
\end{equation}
with $U_{\rm target }$ the targeting gate.
Having ($\beta_j, \varphi_j$), the phase of the laser pulse in Eq.(2) is given, and 
{the modulation amplitude} 
is given by
\bea
A = \frac{|\Delta \tan \beta|}{ \int dx V_{p}(x) \phi_{d}(x)  \phi_{s}(x)}.
\eea
Considering the physical setup, we have $A\in [0, 1]$, this sets a constraint on $\beta$, $\beta \in[0,0.2]$.
In absence of the phase $\varphi$ controlability, the lattice setup would be too restrictive to produce a proper holonomic control unitary with a satisfactory fidelity (Eq.~\eqref{eq:suppfidelity}), which is the reason for us to introduce the control over the phase $\varphi$ in the experiment.  Having both $\beta$ and $\varphi$ under control, we find it is adequate to construct the holonomic $X$, $Y$, $Z$, Hadamard, and $\pi/8$ gates with $M\le 5$. The corresponding control sequences are listed in Table.~\ref{tab:gatesunmodify}.

\begin{table}[htp]
\centering
\caption{Sequences of the holonomic quantum gates before the orbital leakage elimination. 
{Each pulse is applied to the system for one entire period of time $(2\pi/\omega)$.} 
}
\begin{tabular}{ccccccc}
\hline
{gate}                      & { }                           & {pulse 1}   & {pulse 2}   & {pulse 3}   & {pulse 4}   & {pulse 5}\\\hline\hline
\multirow{3}{*}{X-gate}     & {$\omega$/(2$\pi\cdot $kHz)}  & {10.768}    & {10.777}    & {10.769}    & {10.772}    & { }\\
                            & {$A$}                         & {0.6332}    & {0.6505}    & {0.6351}    & {0.6417}    & { }\\
                            & {$\varphi$/rad}               & {1.5377}    & {1.7139}    & {1.5816}    & {1.2258}    & { }\\\hline
\multirow{3}{*}{Y-gate}     & {$\omega$/(2$\pi\cdot $kHz)}  & {10.634}    & {10.736}    & {10.892}    & {10.998}    & { }\\
                            & {$A$}                         & {0.2949}    & {0.5708}    & {0.8311}    & {0.9697}    & { }\\
                            & {$\varphi$/rad}               & {1.0246}    & {0.0415}    & {0.0118}    & {0.0141}    & { }\\\hline
\multirow{3}{*}{Z-gate}     & {$\omega$/(2$\pi\cdot $kHz)}  & {10.892}    & {10.886}    & {10.864}    & {10.889}    & {10.896}\\
                            & {$A$}                         & {0.8318}    & {0.8231}    & {0.7913}    & {0.8274}    & {0.8375}\\
                            & {$\varphi$/rad}               & {0.5454}    & {0.7938}    & {1.5684}    & {2.3366}    & {2.5971}\\\hline
\multirow{3}{*}{Hadamard}   & {$\omega$/(2$\pi\cdot $kHz)}  & {10.815}    & {10.783}    & {10.770}    & {10.796}    & {10.836}\\
                            & {$A$}                         & {0.7147}    & {0.6611}    & {0.6375}    & {0.6828}    & {0.7493}\\
                            & {$\varphi$/rad}               & {2.3780}    & {2.1068}    & {1.5938}    & {1.0489}    & {0.6591}\\
\hline
\multirow{3}{*}{$\pi$/8-gate}   & {$\omega$/(2$\pi\cdot $kHz)}  & {11.023}    & {11.023}    & {11.023}    & {11.023}    & {11.023}\\
                            & {$A$}                         & {0.9987}    & {0.9987}    & {0.9987}    & {0.9987}    & {0.9987}\\
                            & {$\varphi$/rad}               & {1.4075}    & {1.4860}    & {1.5645}    & {1.6431}    & {1.7216}\\
\hline
\end{tabular}
\label{tab:gatesunmodify}
\end{table}

\section{Time-of-flight quantum state tomography}
In our experiment, the momentum distribution $n(\tilde{p})$ (with $\tilde{p}$ the 
momentum along the lattice direction) is obtained with a standard 
tight-of-flight measurement. For the lattice BEC with orbital 
degrees of freedom, the momentum distribution is related to the 
density matrix in Eq. (1) by,
\be
n(\tilde{p}) \propto \sum_{\ell, \nu, \nu'} \delta_{\tilde{p}l, 4\ell\pi} \rho_{\nu \nu'} \tilde{w}_{\nu} (\tilde{p}) \tilde{w}^*_{\nu'} (\tilde{p}),
\label{eq:nq}
\ee
with $\tilde{w}_\nu(\tilde{p})$ the Fourier transform of of the Wannier function of the $\nu$-orbital band.
The momentum distribution has coherent peaks at $\tilde{p} = 4\ell\pi/l$.
In principle, the density matrix can be fully reconstructed from the measured momentum distribution.
But in practice, the higher order momentum peaks with $|\ell|\ge 2$ have rather low signal-to-noise ratio compared to $\ell=0, \pm 1$.
Having the momentum distribution at $\ell=0,\pm 1$ only is not adequate to map out the density matrix.

For quantum state tomography, we let the state to measured evolve in the static lattice for a certain amount of time $t_{\rm evo}$. The time dependence of  the momentum distribution $n(\tilde{p}; t_{\rm evo} )$   is given by Eq.~\eqref{eq:nq} with $\rho_{sd}$ ($\rho_{ds}$) replaced by $\rho_{sd}e^{-i\Delta\times t_{\rm evo}/\hbar}$ ($\rho_{ds}e^{i\Delta\times t_{\rm evo} /\hbar}$). With experimental data for $n(\tilde{p}; t_{\rm evo} )$, the full density matrix $\rho_{\nu \nu'}$ is mapped out by fitting.
{This quantum state tomography scheme for orbital states is analogous to a quench protocol developed for tomography of band insulator topology~\cite{2014_Hauke_PRL}.}

\section{Construction of random unitary gates}

To justify the universality of holonomic gate construction for all single-qubit gates, we randomly sample the SU(2) unitary gates, and obtain its fidelity in atom-orbital gate realization through our multi-orbital simulation method, which has been found to model the experiment quantitatively. For an SU(2) unitary operation, parameterized as $e^{i \sigma ({\theta, \phi}) \beta /2}$, with $\beta$ the rotation angle on the Bloch sphere, and $\theta$ and $\phi$ defining the rotation
axis---$\sigma(\theta, \phi) = \cos(\theta) \sigma_z + \sin(\theta)\left[ \cos(\phi) \sigma_x + \sin(\phi) \sigma_y \right]$.
We randomly sample the angles according to a uniform distribution, $\theta \in [0, \pi)$, $\phi \in[0, 2\pi)$, and $\beta \in [0, 2\pi)$. Through the orbital leakage error elimination protocol, we confirm all these generated random unitary gates can be performed with our atom-orbital {qubit}, with final state fidelities all above $98\%$ (Figure~\ref{fig:random}).

\begin{figure}
\centering
\includegraphics[width=0.6\textwidth]{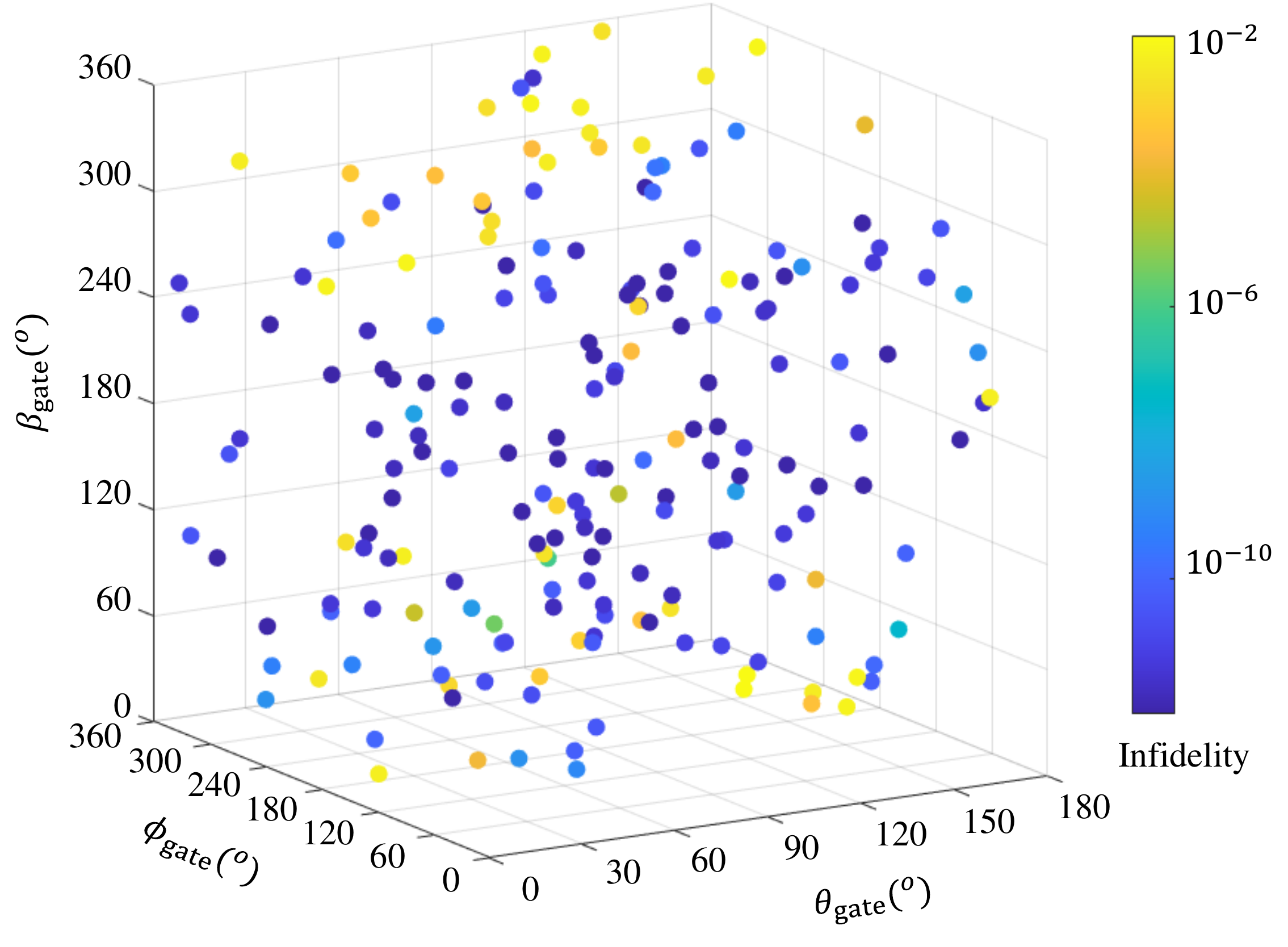}
\caption{Final state fidelities of $100$ random SU(2) rotations. We randomly sample rotation axis (represented by the polar angle $\theta_{gate}$ and the azimuth angle $\phi_{gate}$) and the rotation angle $\beta_{gate}$.
For each of these quantum gates, we theoretically design the holonomic quantum control sequence and calculate the corresponding final state fidelity using our multi-orbital numerical simulations. All the gates are achieved with the pulse step $M\le 5$. All fidelities are above $98\%$.
}
\label{fig:random}
\end{figure}

\begin{figure}[htp]
\centering
\includegraphics[width=1\textwidth]{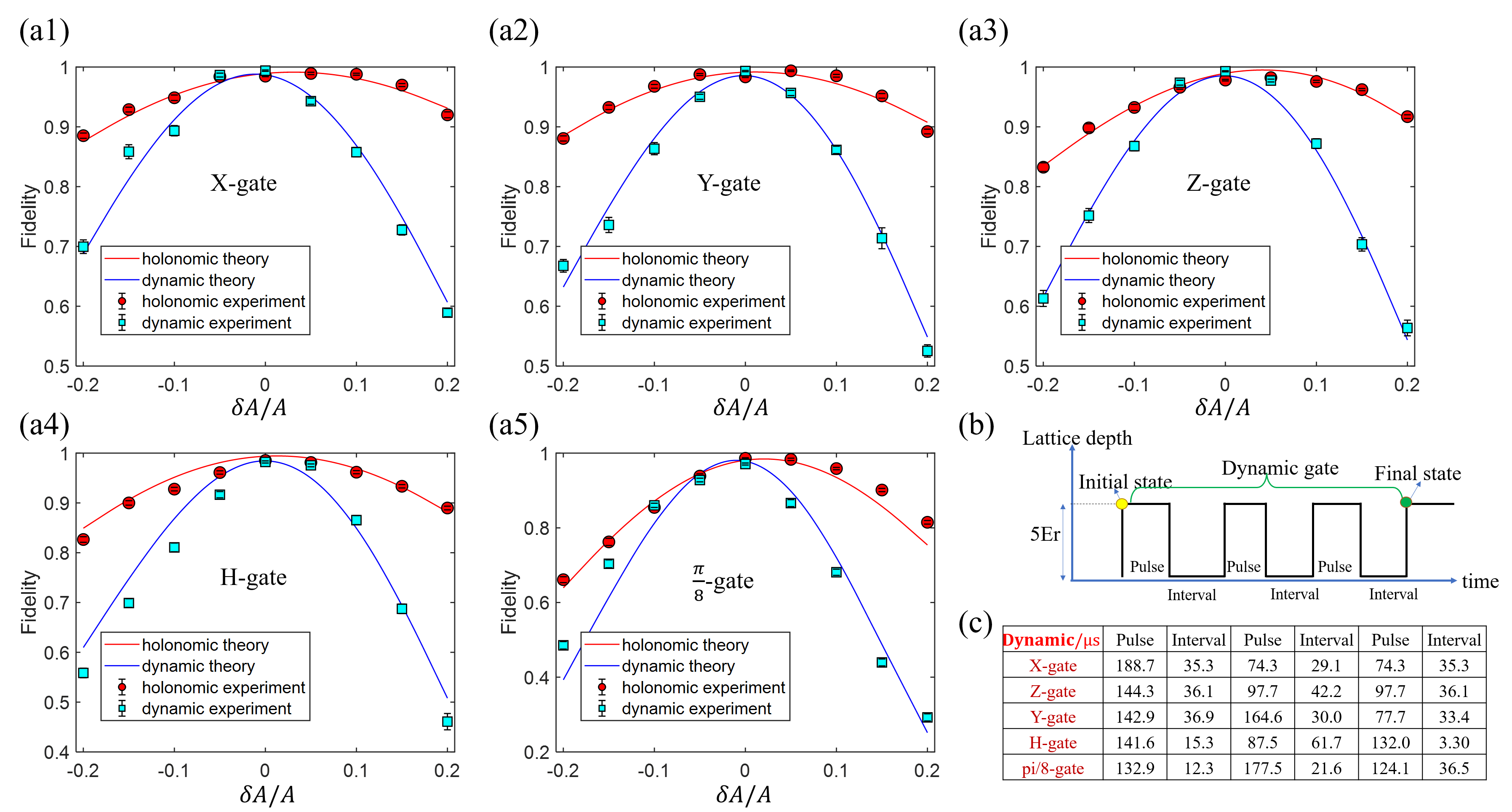}
\caption{{Comparison between holonomic gates and non-holonomic dynamic gates. (a) Gate robustness against laser fluctuation noise. 
The ratio ($\delta A/A$) represents the relative fluctuation strength in the laser intensity. 
{DC-noise is applied here to account for long-term drift of laser intensities that may potentially happens in experiments.}  
The red points (blue squares) represent the experimental fidelity of the holonomic gates (dynamic gates). The fidelity is obtained by the QPT method. The red lines (blue lines) are the theoretical results (simulated with the multi-orbital 
numerical method) for the fidelity of the holonomic gates (dynamic gates). (a1) to (a5) correspond to the X-, Y-, Z-, Hadamard-, and $\pi/8$- gate, respectively.
(b) Time sequence of the non-holonomic dynamic gate. The dynamic gate is constructed by a series of square wave pulses. The duration and interval for each pulse are designed and optimized to realize different gates.  The pulse sequence details are provided in (c).}}
\label{fig:gaterobust}
\end{figure}

{
\section{Robustness against laser fluctuations}
Having robustness against noise in the quantum control protocol is critical to reaching high quality quantum information processing. We characterize  the robustness of the holonomic quantum gates by deliberately introducing a slight deviation on the optical lattice potential amplitude in Eq. (2) with an amount of $\delta A$.
The results of the robustness study are provided in Fig.~\ref{fig:gaterobust}. As the lattice modulation amplitude varies up to $10\%$, we find the average experimental fidelity of the holonomic gates remains above $90\%$. 
The experimental results have a quantitative agreement with the multi-orbital numerical simulation (Fig.~\ref{fig:gaterobust}(a)). 
}

{
In order to further illustrate the advantage of the holonomic gates, we also construct a non-holonomic dynamic gate, and show their comparison in Fig.~\ref{fig:gaterobust}(a).
As Fig.~\ref{fig:gaterobust}(b) shows, the non-holonomic dynamic gate is constructed by a series of square wave pulses for the lattice depth, and its dynamic phase is not cancelled. The time duration and interval in each pulse are optimized to reach  the targeting gate operation.  
The specific time sequences are shown in Fig.~\ref{fig:gaterobust}(c), and the total time is close to the holonomic gate for a fair comparison.
Both gates can achieve high fidelity without the laser intensity noise (when $\delta A/A=0$). 
However, by introducing laser intensity noise with $\delta A/A \neq 0$, we find that the holonomic gates exhibit better resistance against fluctuations. 
For example, the QPT fidelity of the holonomic X-gate is above 92\% when the fluctuation $\delta A/A$ reaches  $20\%$. In sharp contrast, the fidelity of the non-holonomic dynamic X-gate is below $60\%$. 
The noise resilience of the holonomic quantum gates on orbital qubits is thus confirmed in our experiment.
}

{In principle, imperfect control of the pulse durations may also affect the gate quality. But in optical lattice experiments, the control precision of pulse duration is typically 
maintained below the level of  $10^{-4}$, which could only cause fidelity decreasing by about  $0.01\%$ through our numerical simulation.  
}

 \begin{figure*}
\centering
\includegraphics[width=0.7\textwidth]{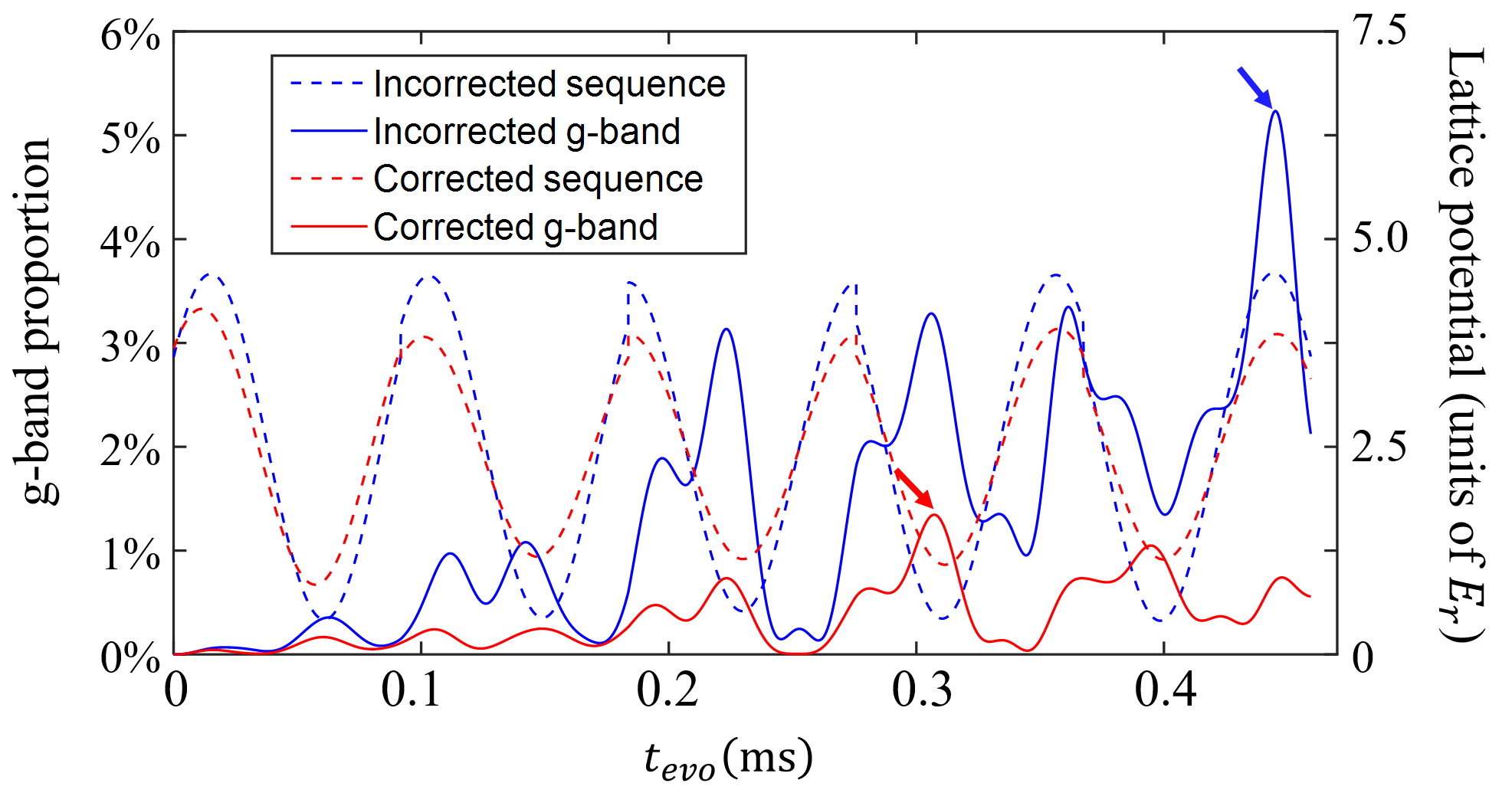}
\caption{
Leakage to the $g$-band. 
The red dashed and  blue dashed lines represent the modulation sequences of the  Z gate with and without our orbital leakage  elimination scheme, respectively. 
The red solid and blue solid  lines illustrate the proportion of leakage to the g-band during the Z gate process with and without the orbital leakage elimination.  
The red and  blue arrows mark the maximum proportion of the $g$-band for the two cases. 
Here we choose  the initial state to be $|1\rangle$.}
\label{fig:gband}
\end{figure*}

\section{Orbital leakage elimination}
In the experiment, we observe that the orbital leakage error strongly compromises the orbital qubit gate fidelity. Although we try to avoid the transitions to the unwanted orbitals by energy suppression and symmetry, the leakage error is still present and harmful. As an example, we show the process of Z-gate operation on the initial state $|1\rangle$ in Fig.~\ref{fig:gband}. Through our multi-orbital numerical simulation, we find that $g$-band proportion reaches above $5\%$, and the final state fidelity is about $71\%$.  

In order to resolve the orbital leakage problem, we develop an orbital leakage elimination technique, to be described in the next paragraph. With the orbital leakage elimination technique, the proportion of the g-band drops down below $1.3\%$ (Fig.~\ref{fig:gband}), and the fidelity reaches above $99\%$ for the Z-gate operation, in our numerical simulation. The orbital leakage elimination technique is implemented in our experiment. This technique is the key for us to reach the high gate fidelities as reported in the main text.

Here we describe the details of orbital leakage elimination technique. 
For a target holonomic quantum gate, we first obtain the parameter sequence ($\theta_j, \varphi_j$) by maximizing the idealized gate fidelity (the index $j$ labels the lattice modulation periods). Then at each lattice modulation period, the quantum state evolution $U^\star (t)$ ($t\in [0, T]$) is fully specified by $\theta_j$ and $\varphi_j$.
The orbital leakage elimination scheme is to search for a most suitable choice of lattice modulation pulse, i.e., ($\omega$, $A$, $\varphi$), which gives a quantum evolution that best approximates $U^\star (t)$. In order to reach this solution, we propose a  loss function
\bea
L(\Theta)&=&
\int dt \left\{ \Sigma_{nm\in\{S,D\}}|U^{\star}_{nm} (t)-U_{nm}(t; \Theta)|^2  \right. +  \\
&&\left. \Sigma_{n\in\{S,D\},m\notin\{S,D\}} |U_{nm}(t; \Theta)|^2+|U_{mn}(t; \Theta)|^2 \right\}, \nn
\eea
where $\Theta$ represents $( \omega, A, \varphi)$, and $U(t; \Theta)$ the quantum evolution matrix incorporating all orbitals in our calculation.
Through minimization, we obtain a most suitable $\Theta$, with which the leakage error caused by the unwanted orbitals is minimized.
The control sequences designed with the orbital leakage error elimination protocol for $X$, $Y$, $Z$, Hadamard, and $\pi/8$ gates are listed below as Table.~\ref{tab:gatesequence}.

\begin{table}[htp]
\centering
\scalebox{0.93}{
\begin{tabular}{ccccccc}
\hline
{gate}                      & { }                           & {pulse 1}   & {pulse 2}   & {pulse 3}   & {pulse 4}   & {pulse 5}\\\hline\hline
\multirow{3}{*}{X-gate}     & {$\omega$/(2$\pi\cdot $kHz)}  & {10.768}    & {10.777}    & {10.769}    & {10.772}    & { }\\
                            & {$A$}                         & {0.4749}    & {0.4555}    & {0.5626}    & {0.4765}    & { }\\
                            & {$\varphi$/rad}               & {1.6142}    & {1.6383}    & {1.5791}    & {1.2129}    & { }\\\hline
\multirow{3}{*}{Y-gate}     & {$\omega$/(2$\pi\cdot $kHz)}  & {10.634}    & {10.736}    & {10.892}    & {10.998}    & { }\\
                            & {$A$}                         & {0.2027}    & {0.3952}    & {0.6141}    & {0.7747}    & { }\\
                            & {$\varphi$/rad}               & {0.6247}    & {0.2082}    & {0.1576}    & {0.1429}    & { }\\\hline
\multirow{3}{*}{Z-gate}     & {$\omega$/(2$\pi\cdot $kHz)}  & {10.892}    & {10.886}    & {10.864}    & {10.889}    & {10.896}\\
                            & {$A$}                         & {0.6639}    & {0.5293}    & {0.5407}    & {0.5680}    & {0.5428}\\
                            & {$\varphi$/rad}               & {0.7953}    & {0.9442}    & {1.5551}    & {2.2610}    & {2.4972}\\\hline
\multirow{3}{*}{Hadamard}   & {$\omega$/(2$\pi\cdot $kHz)}  & {10.815}    & {10.783}    & {10.770}    & {10.796}    & {10.836}\\
                            & {$A$}                         & {0.5153}    & {0.5106}    & {0.5445}    & {0.5281}    & {0.4986}\\
                            & {$\varphi$/rad}               & {2.3395}    & {2.0068}    & {1.5531}    & {1.1157}    & {0.8583}\\
\hline
\multirow{3}{*}{$\pi$/8-gate}   & {$\omega$/(2$\pi\cdot $kHz)}  & {11.023}    & {11.023}    & {11.023}    & {11.023}    & {11.023}\\
                            & {$A$}                         & {0.8012}    & {0.8030}    & {0.7970}    & {0.8176}    & {0.6218}\\
                            & {$\varphi$/rad}               & {1.5058}    & {1.4766}    & {1.5636}    & {1.6471}    & {1.6892}\\
\hline
\end{tabular}}
\caption{The control sequences of the holonomic quantum gates with the orbital leakage elimination protocol. Each pulse is applied to the system for one entire period of time $(2\pi/\omega)$.
}
\label{tab:gatesequence}
\end{table}

 \begin{figure*}
\centering
\includegraphics[width=1\textwidth]{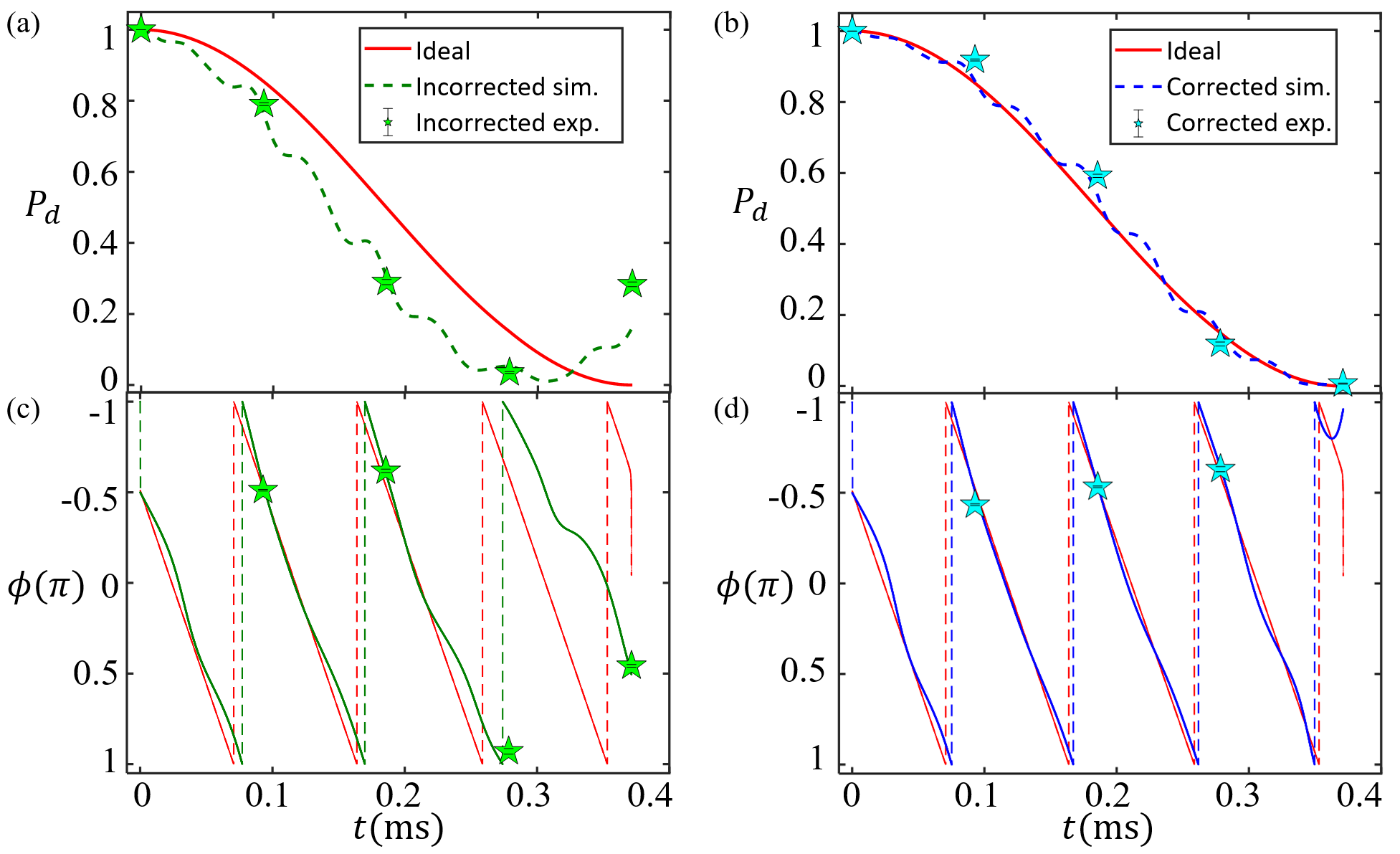}
\caption{Comparison between experimental observation with numerical 
simulation for the quantum state evolution during the holonomic quantum control.
The dynamical process corresponds to a holonomic $X$-gate on the initial state, 
$|0\rangle$ ($d$-orbital state). (a) and (b) show the time evolution of the  proportion of the $d$-band population, $P_d$.
(c) and (d) show the time evolution of the phase difference between $|0\rangle$ ($d$-orbital state) and $|1\rangle$ ($s$-orbital state).
 The red lines represent the ideal state evolution of the holonomic control.
 The blue and green stars correspond to the experimental results with and without the orbital leakage elimination protocol.
 The green and blue lines are the corresponding simulation results.}
\label{fig:numexp}
\end{figure*} 

\section{Multi-orbital simulation method}
In the experiment, we find the discrepancy in the simplified two-orbital model in Eq.~(4)---the simulated results have significant disagreement with the experiment. We thus develop a multi-orbital simulation method, which treats the non-interacting quantum dynamics exactly.  Having atoms confined in a dynamical potential $V(x,t) = V(t) \cos^2 (Kx)$, the Bose-Einstein condensate state evolves in a subspace spanned by plane wave states $|q \rangle$, with its wave function $e^{2iqKx}$. In this subspace, the kinetic part of the Hamiltonian is diagonal given by
\be
H_{K, qq'} = \frac{\hbar^2}{2M}  (2qK)^2 \delta_{qq'}.
\ee
The potential term is off-diagonal given by
\be
H_{P, qq'} = \frac{1}{2} V(t) \times \left( \delta_{q,q'+1} + \delta_{q, q'-1}  \right).
\ee
The quantum state  evolution is described by a time-dependent column vector $\psi(t)$, according to
\be
\psi(t+\delta t) = e^{-i H_K \delta t/2} e^{-iH_P (t+\delta t/2) \delta t} e^{-i H_K \delta t/2} \psi(t) + {\cal O} (\delta t^3) .
\ee
Since $H_K$ is diagonal, the multiplication of $e^{-i H_K \delta t}$ on a vector can be efficiently implemented as multiplication of an element-wise phase. The matrix $H_P$ is off-diagonal, but can be diagonalized by a Fourier transformation. The matrix multiplication of $e^{-iH_K \delta t}$ can  be efficiently performed by combining with a Fast Fourier transformation algorithm. Choosing a high-momentum cutoff $|q|<q_{\rm max}$, the computation complexity of this multi-orbital simulation method is ${\cal O} (q_{\rm max} \log q _{\rm max} )$. The simulated results have very good agreement with our experimental measurements (Fig.~\ref{fig:numexp}). With numerical simulation, we find the orbital leakage error strongly damages the quantum gate operations. With our orbital leakage elimination protocol, the time evolution taking the corrected control pulses approaches to the ideal holonomic quantum dynamics.

\section{Construction of multiple qubits and two-qubit operations}
In our experiment, we have performed single qubit control based on the orbital state of a BEC. To engineer scalable quantum circuits with the orbital qubit setup, we need to introduce single-site controllablility, which is experimentally accessible with  quantum microscope techniques~\cite{wang2016single,2020_Li_NPJ}. To selectively control the orbital transition on one targeting site, say at $x_0$, we propose to maintain the optical potential approximately homogeneous  elsewhere, and tune the gap between $s$ and $d$ orbitals at site $x_0$ far off-resonant from other sites by manipulating the local potential using quantum microscope techniques.  The overall optical potential is now denoted as $\tilde{V}(x)$, and the gap between $s$ and $d$ orbitals at the targeting site is $\tilde{\Delta}_0$. The orbital transition at $x_0$ can then be controlled following the same way as demonstrated experimentally in this work, namely by introducing a modulating lattice potential, 
\be 
\Delta \tilde{V}(x, t)  = A \sin (\omega t + \varphi) \tilde{V} (x) 
\ee 
The modulation frequency $\omega$ should be chosen to match the gap $\tilde{\Delta}_0$. Working in the limit of one particle per lattice site, the interaction effects during the control process of the single-qubit rotation are negligible. The integration of quantum microscope techniques to the present holonomic orbital qubit control would make an orbital-based quantum computing platform with single-qubit addressibility. 

The two-qubit gate can be achieved by adapting the atom collision scheme~\cite{2008_Porto_PRA,2019_Sherson_PRA,2020_Rey_AQT,2021_Zwierlein_arXiv}. Considering two nearby sites, denoted as $a$ and $b$, the orbital states are $|s_a s_b\rangle$, $|d_a d_b\rangle$, $|s_a d_b\rangle$, and $|d_a s_b\rangle$. The local interactions projected to this subspace give a vanishing contribution for each site only contains one particle. The leading order interaction that contributes to the dynamics in the two-qubit computation subspace is thus between neighboring sites. 
Given the presence of a large gap between $s$ and $d$ orbitals, the interaction processes that do not respect the conservation of single-particle energy are negligible. 
The relevant interactions between neighboring sites are then 
\be 
H_{\rm int,nn} = U_1 \hat{s}_a^\dag \hat{s}_b^\dag \hat{s}_b \hat{s}_a  + U_2 \hat{d}_a ^\dag \hat{d}_b ^\dag \hat{d}_b \hat{d}_a 
+ U_3 \left( \hat{s}_a ^\dag \hat{d}_b ^\dag \hat{d}_b \hat{s}_a + \hat{d}_a^\dag \hat{s}_b^\dag \hat{s}_b \hat{d}_a  \right)  
+ U_4 \left( \hat{s}_a^\dag \hat{d}_b^\dag \hat{s}_b \hat{d}_a + \hat{d}_a^\dag \hat{s}_b^\dag \hat{d}_b \hat{s}_a   \right)   
\ee  
where we have introduced orbital annihilation operators $\hat{s}_a$, $\hat{s}_b$, $\hat{d}_a$, $\hat{d}_b$ on the two sites. For the orthogonality between different wannier functions on each site, $U_4$ is much smaller than $U_3$, and can be neglected. The interactions $U_1$ and $U_2$ only contribute to the diagonal terms  in the two-qubit computation subspace. The $U_3$ process is more nontrivial, causing a swap operation between $|s_a d_b \rangle$ and $|d_a s_b \rangle$. The swap operation is completed with the interaction turned on for a time duration 
\be 
\tau_{\rm swap} = \pi\hbar /(2 U_3 ).
\ee  
A $\sqrt{\rm swap}$ gate is then reached by letting atoms interact for  a time duration 
\be 
\tau_{\sqrt{\rm swap}} = \tau_{\rm swap}/2.
\ee 
This makes an  entangling two-qubit gate, which is sufficient to make a universal gate set by combining with single-qubit operations~\cite{nielsen2002quantum}. 
In order to make an efficient $\sqrt{\rm swap}$ gate, it is required to move the two sites closer  than the lattice spacing by adiabatic time evolution.   This  requires quantum microscope or sublattice techniques.


\end{widetext} 

\end{document}